\documentclass{mp4}
\usepackage{amsmath,amsfonts,amssymb}
\usepackage{url}
\DeclareMathOperator{\sgn}{sgn}

\begin{document}

\markboth{K. Wawrzyniak and W. Wislicki}
{Phenomenology of minority games in efficient regime}

%
%

\title{PHENOMENOLOGY OF MINORITY GAMES IN EFFICIENT REGIME 
}

\author{KAROL WAWRZYNIAK}

\address{Interdisciplinary Centre for Mathematical and Computational Modelling, University of Warsaw, Pawi\'nskiego 5A, PL-02-106 Warszawa, Poland\\
K.Wawrzyniak@icm.edu.pl}

\author{WOJCIECH WISLICKI}

\address{A. So\l tan Institute for Nuclear Studies, Ho\.za 69, PL-00-681 Warszawa, Poland\\
wislicki@fuw.edu.pl}

\maketitle

\begin{history}
\end{history}

\begin{abstract}
We present a comprehensive study of utility function of the minority game in its efficient regime.
We develop an effective description of state of the game.
For the payoff function $g(x)=\sgn (x)$ we explicitly represent the game as the Markov process and prove the finitness of number of states.
We also demonstrate boundedness of the utility function.
Using these facts we can explain all interesting observable features of the aggregated demand: appearance of strong fluctuations, their periodicity and existence of preferred levels.
For another payoff, $g(x)=x$, the number of states is still finite and utility remains bounded but the number of states cannot be reduced and probabilities of states are not calculated.
However, using properties of the utility and analysing the game in terms of de~Bruijn graphs, we can also explain distinct peaks of demand and their frequencies.
\end{abstract}

\keywords{Minority game, adaptive system, Markov process, de Bruijn graph}

\section{Introduction}

Minority game (MG) was designed \cite{arthur} as a microscopic model of adaptive behaviour observed in multi-agent systems.
The MG is a typical bottom-up construct and therefore usual definitions of the game first specify rules of behaviour for individuals.
Then, piecing together microscopic variables, one defines higher-order quantities characterizing grander systems.
In some cases, however, other constructs are also possible, e.g. functions of state like score functions can be attributed to groups of agents without specifying agents individually (cf. ref. \cite{jeffries_1}).
Despite simplicity of basic rules of taking decisions by agents, adaptive abilities and phenomenology of populations playing MGs appear to be surprisingly interesting and their properties are non-trivial \cite{challet_1}.
Special studies were devoted to understanding of such functions like aggregated demand, market volatility, market occupancy etc.
It was shown \cite{challet_2,savit} that the MG exhibits different modes of behaviour, depending on the game parameters: the random, cooperation and herd.
The latter case is characterized by small strategy space compared to the overall number of agents.
Following authors of ref. \cite{decara} we prefer to call this regime {\it efficient}, because all players have all available information at their disposal.
Our study of this regime is motivated by interesting phenomenology observed in numerical simulations and lack of satisfactory interpretations of them.
For example, the aggregated demand exhibits large-amplitude oscillations \cite{savit} and periodicity in time \cite{decara,zheng}.
The crowd-anticrowd theory \cite{hart_1,hart_2} presents acceptable explanation for oscillations but fails to deal with the periodicity.
This issue was treated by the authors of ref. \cite{zheng} and, more fruitfully, ref. \cite{jeffries_1}.
The authors of ref.~\cite{jeffries_1} introduced the concept of the state of the MG but limit their analysis to the reduced strategy space.

In our previous work \cite{wawrzyniak} we found, in different context, that the crucial role in explanation of observable behaviour in the MG is played by the utility function.
Therefore in this paper we further exploit the utility to study phenomenology of MGs in their efficient regime.
We find that the utility is bounded and the number of states is finite, and prove these facts for the payoff function $g(x)=\sgn(x)$.
We can represent the game as a Markov process and we can substantially reduce the number of states and calculate their probabilities.
Then such interesting features of demand like its strong inhomogeneity and presence of patterns in time can be easily interpreted.
For other payoff functions, e.g. $g(x)=x$, the number of states cannot be reduced and distribution of utility remains irregular. 
In this case we cannot explicitly calculate probabilities of states.
However, using the same general properties of the utility and representing the game as paths on de~Bruijn diagrams, we can also explain strong fluctuations of demand and calculate their frequency.

\section{Formal definition of the minority game}

At each time step $t$, the $n$-th agent out of $N$ $(n=1,\ldots,N)$ takes an action $a_{\alpha_n(t)}$ according to some strategy $\alpha_n(t)$.
The action $a_{\alpha_n(t)}$ takes either of two values: $-1$ or $+1$.
An aggregated demand is defined 
\begin{eqnarray}
A(t)=\sum_{n=1}^{N}a_{\alpha_n^\prime(t)},
\label{eq11}
\end{eqnarray}
where $\alpha_n^\prime$ refers to the action according to the best strategy, as defined in eq.~(\ref{eq13}) below.
Such defined $A(t)$ is the difference between numbers of agents who choose the $+1$ and $-1$ actions.
Agents do not know each other's actions but $A(t)$ is known to all agents.
The minority action $a^\ast(t)$ is determined from $A(t)$
\begin{eqnarray}
a^\ast(t)=-\sgn A(t).
\label{eq12}
\end{eqnarray}
Each agent's memory is limited to $m$ most recent winning, i.e. minority, decisions.
Each agent has the same number $S\ge 2$ of devices, called strategies, used to predict the next minority action $a^\ast(t+1)$.
The $s$-th strategy of the $n$-th agent, $\alpha_n^s$ $(s=1,\ldots,S)$, is a function mapping the sequence $\mu$ of the last $m$ winning decisions to this agent's action $a_{\alpha_n^s}$.
Since there is $P=2^m$ possible realizations of $\mu$, there is $2^P$ possible strategies.
At the beginning of the game each agent randomly draws $S$ strategies, according to a given distribution function $\rho(n):n\rightarrow \Delta_n$, where $\Delta_n$ is a set consisting of $S$ strategies for the $n$-th agent.

Each strategy $\alpha_n^s$, belonging to any of sets $\Delta_n$, is given a real-valued function $U_{\alpha_n^s}$ which quantifies the utility of the strategy: the more preferable strategy, the higher utility it has.
Strategies with higher utilities are more likely chosen by agents.

There are various choice policies.
In the popular {\it greedy policy} each agent selects the strategy of the highest utility
\begin{eqnarray}
\alpha_n^\prime(t)=\arg \max_{s:\,\alpha_n^s \in \Delta_n} U_{\alpha_n^s}(t).
\label{eq13}
\end{eqnarray}
If there are two or more strategies with the highest utility then one of them is chosen randomly.
The highest-utility strategy (\ref{eq13}) used by the agent is called the {\it active strategy}, in contrast to {\it passive strategies}, unused at given moment.
However, at any time all agents evaluate all their strategies, the active and passive ones.
Each strategy $\alpha_n^s$ is given the {\it payoff} depending on its action $a_{\alpha_n^s}$
\begin{eqnarray}
R_{\alpha_n^s}(t)=-a_{\alpha_n^s}(t)\,g[A(t)],
\label{eq14}
\end{eqnarray}
where $g$ is an odd {\it payoff function}, e.g. the steplike $g(x)=\sgn(x)$ \cite{challet_2}, proportional $g(x)=x$ or scaled proportional  $g(x)=x/N$.
The learning process corresponds to updating the utility for each strategy
\begin{eqnarray}
U_{\alpha_n^s}(t+1)=U_{\alpha_n^s}(t)+R_{\alpha_n^s}(t),
\label{eq15}
\end{eqnarray}
such that every agent knows how good its strategies are. 

\section{Phenomenology}

In order to examine MGs in the efficient regime, we performed a series of numerical simulations with different combinations of game parameters, and chosen three most representative cases: $(m,N)=(1,401),(2,1601),(5,1601)$, all with the number of strategies per agent $S=2$.
All three games are in the efficient mode.
In the first two cases the condition $NS\gg 2^P$ is fulfilled.
In the third one it is not met and consequences of this fact will become clear later in the text.
In all three experiments the full strategy space is used.

The effective mode is often called {\it symmetric phase} in the literature (cf. e.g. ref. \cite{challet_3}) which means that both actions are taken by the minority agents with the same frequency.

\begin{figure}[t]
\begin{center}
\begin{tabular}{ccc}
\includegraphics[scale=.27]{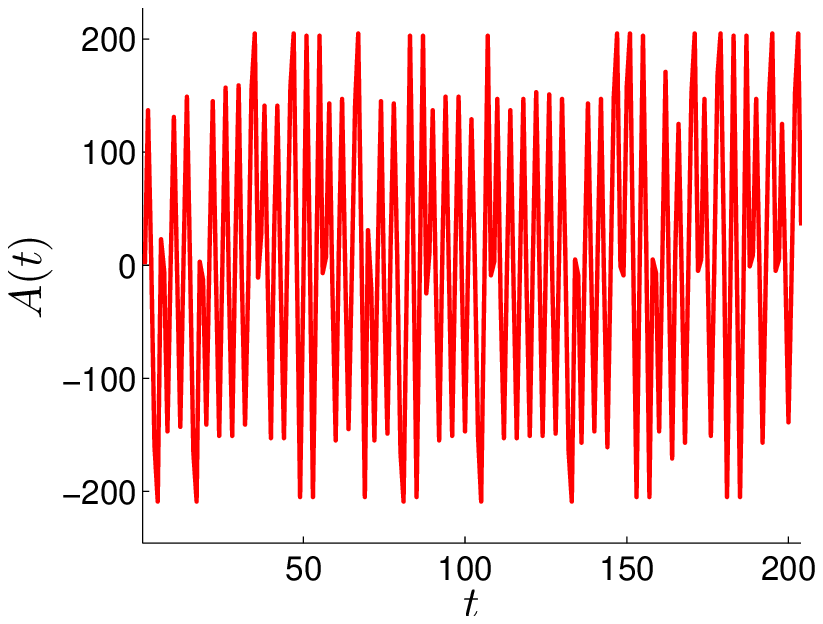} & \includegraphics[scale=.27]{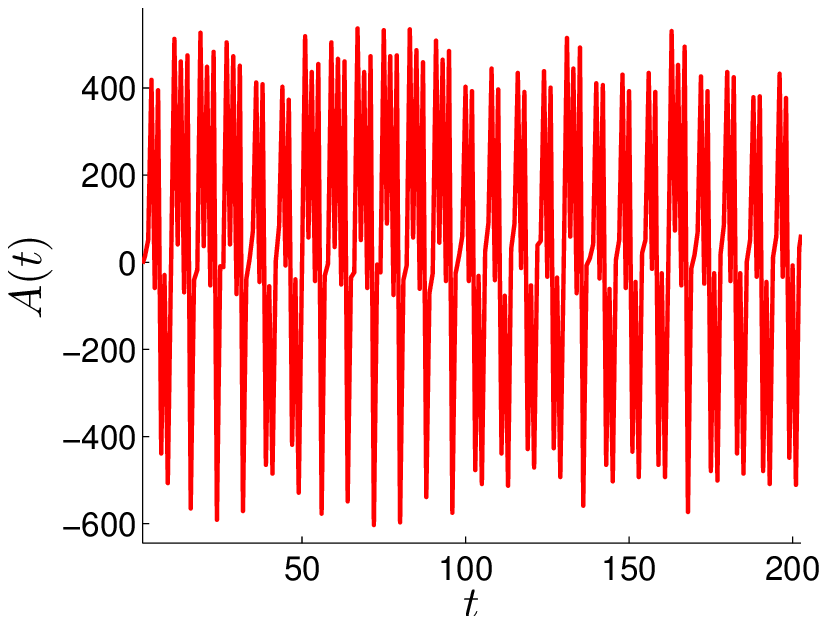} & \includegraphics[scale=.27]{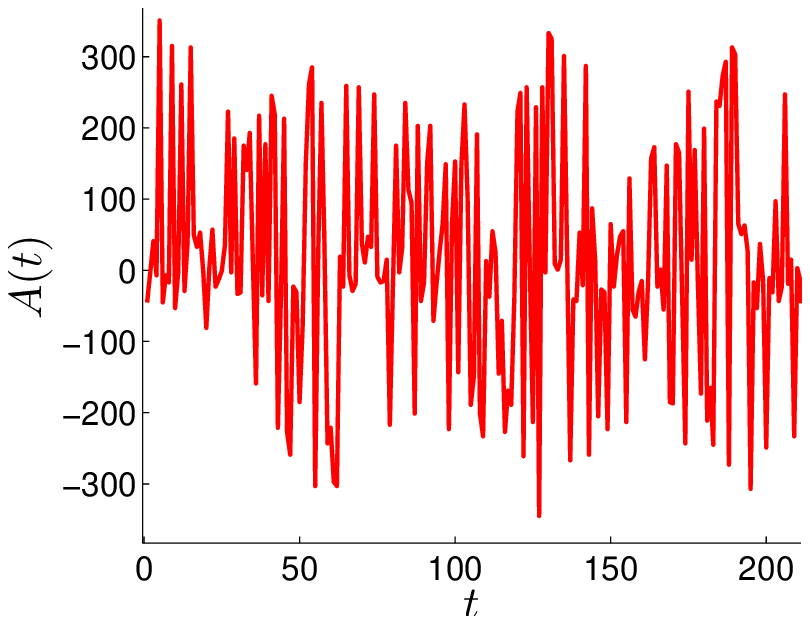}
\end{tabular}
\end{center}
\vspace*{8pt}
\caption{\label{fig31} Time evolution of the aggregated demand $A(t)$ for three combinations of the population size $N$ and agent memory $m$: $N=401$, $m=1$ (left), $N=1601$, $m=2$ (middle) and $N=1601$, $m=5$ (right). Simulations were done for $S=2$ and $g(x)=\sgn (x)$. Preferred values of $A$ are visible for all three games.}
\end{figure}

\begin{figure}[t]
\begin{center}
\begin{tabular}{ccc}
\includegraphics[scale=.27]{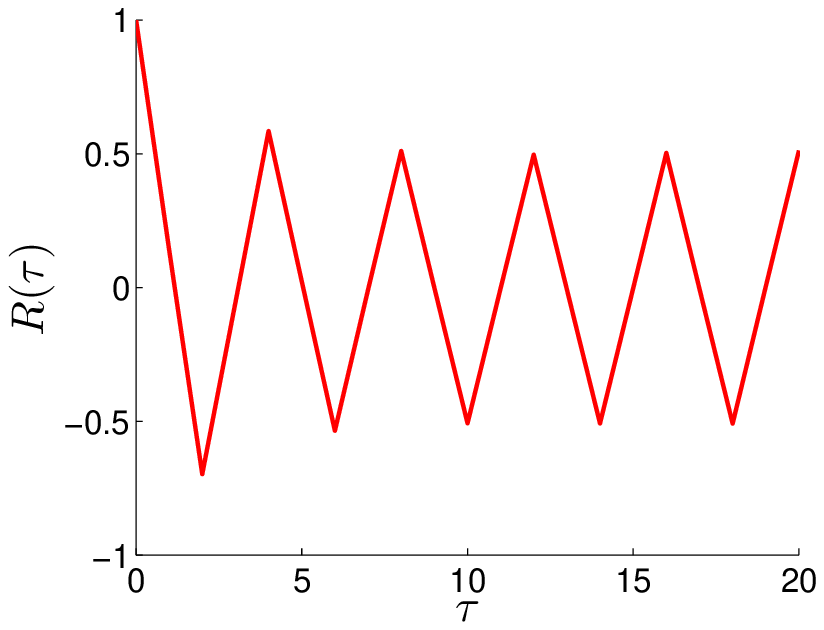} & \includegraphics[scale=.27]{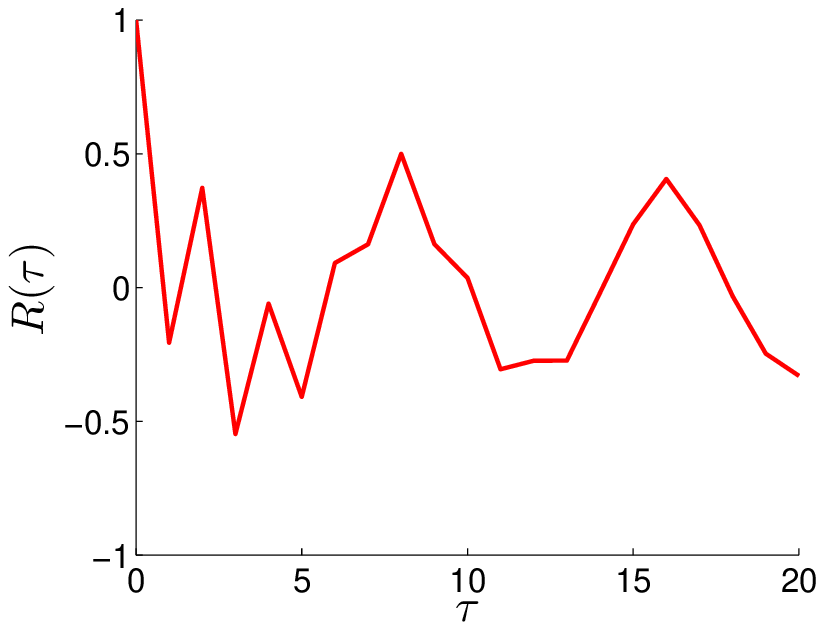} & \includegraphics[scale=.27]{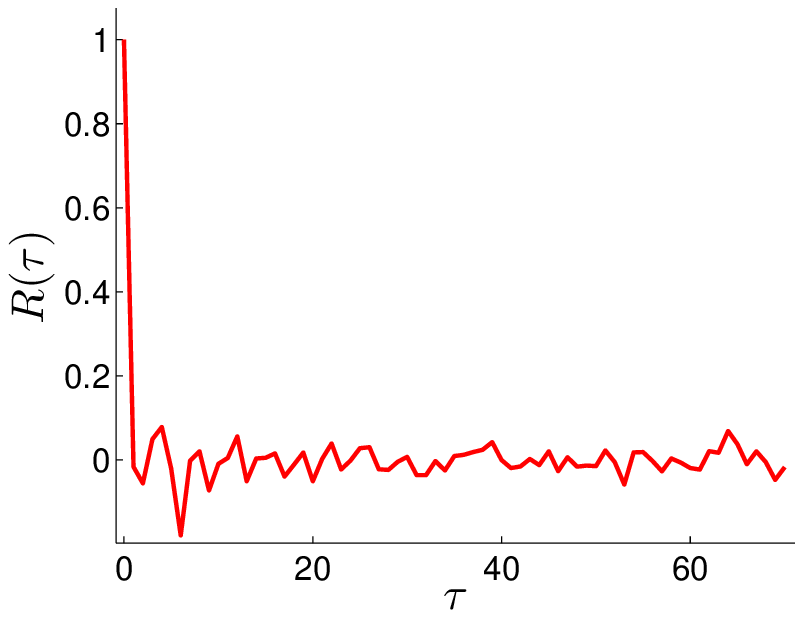}
\end{tabular}
\end{center}
\vspace*{8pt}
\caption{\label{fig32} Autocorrelation function $R(\tau)$ for three combinations of the population size $N$ and agent memory $m$: $N=401$, $m=1$ (left), $N=1601$, $m=2$ (middle) and $N=1601$, $m=5$ (right). Simulations were done for $S=2$ and $g(x)=\sgn (x)$. The highest values of $R$ are for $\tau = 2\cdot 2^m$, except for $\tau=0$, for all games fulfilling the $NS\gg 2^P$ condition.}
\end{figure}

\begin{figure}[t]
\begin{center}
\begin{tabular}{ccc}
\hspace{-8pt} \includegraphics[scale=.2]{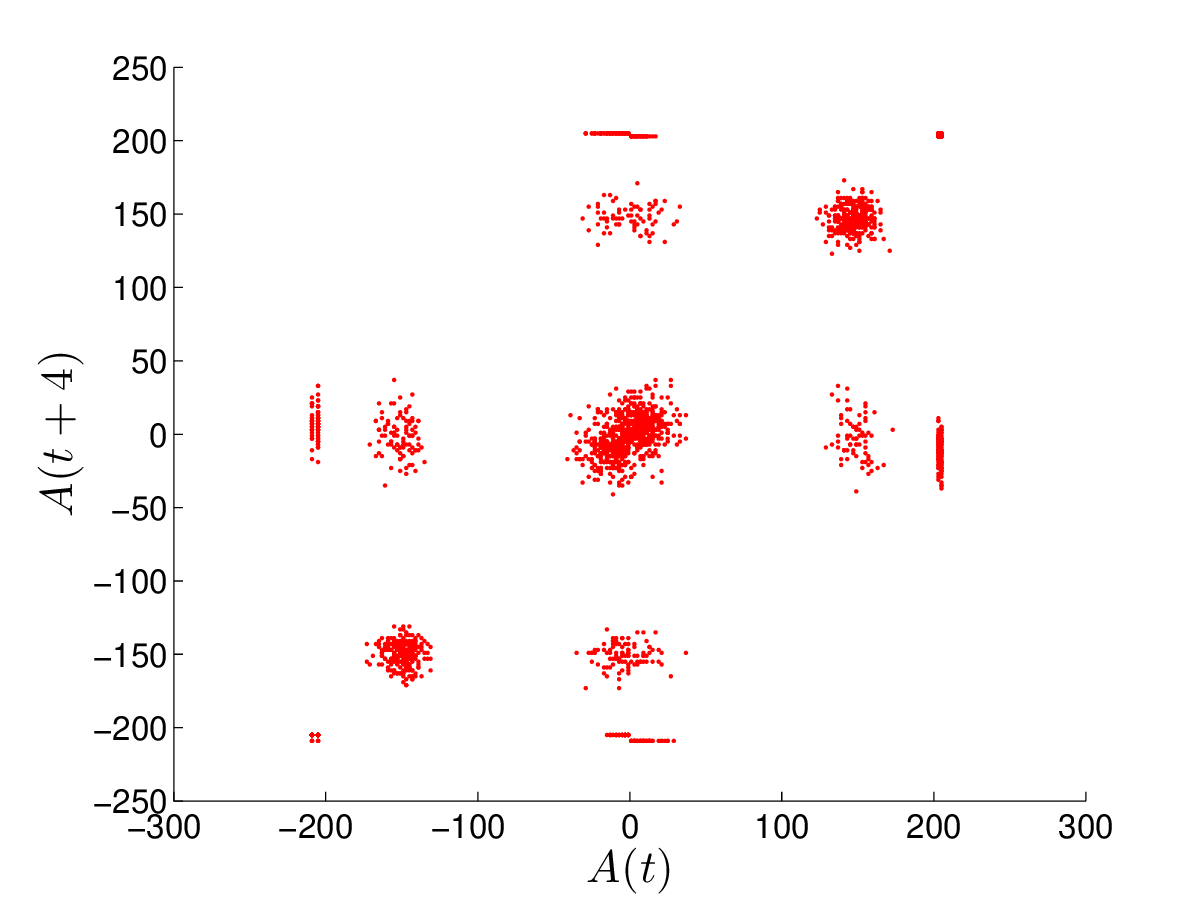} & \hspace{-8pt} \includegraphics[scale=.2]{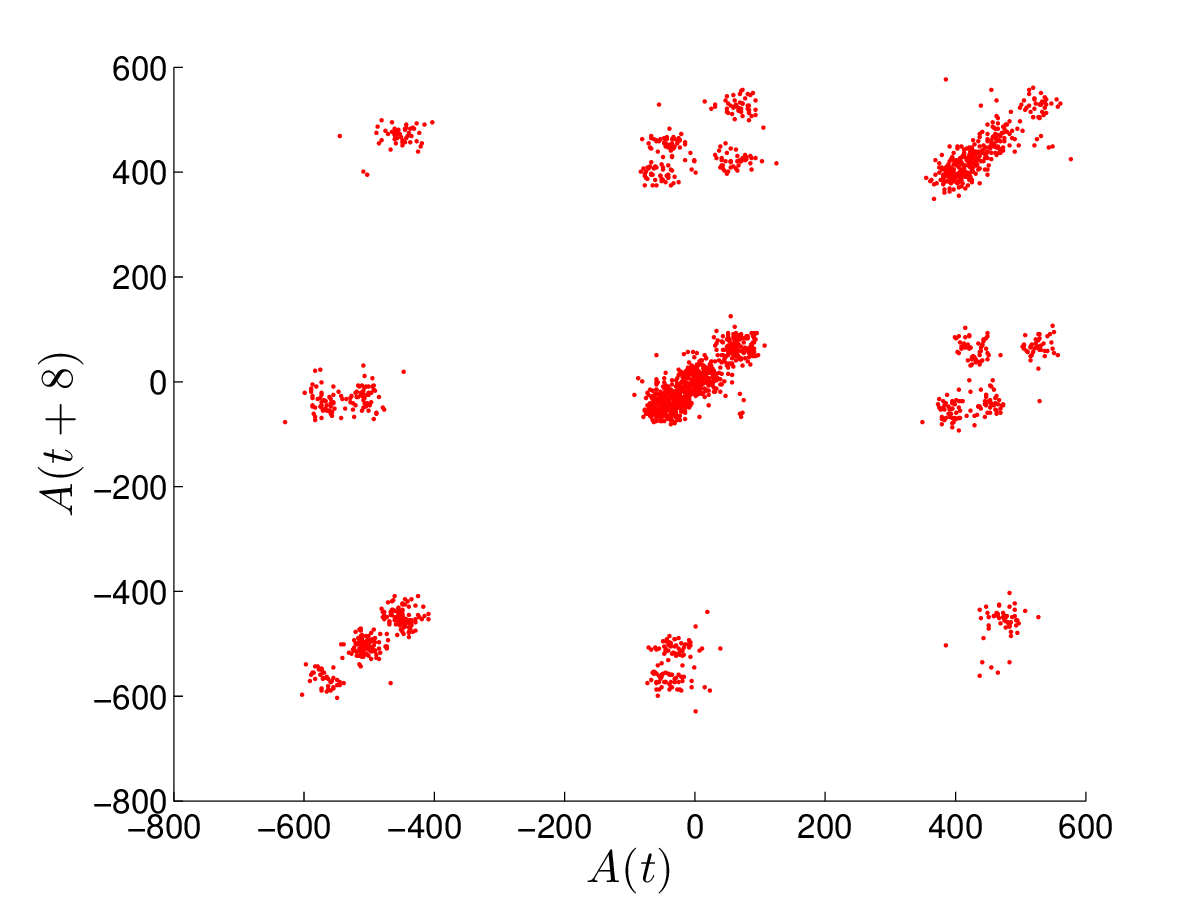} & \hspace{-8pt} \includegraphics[scale=.3]{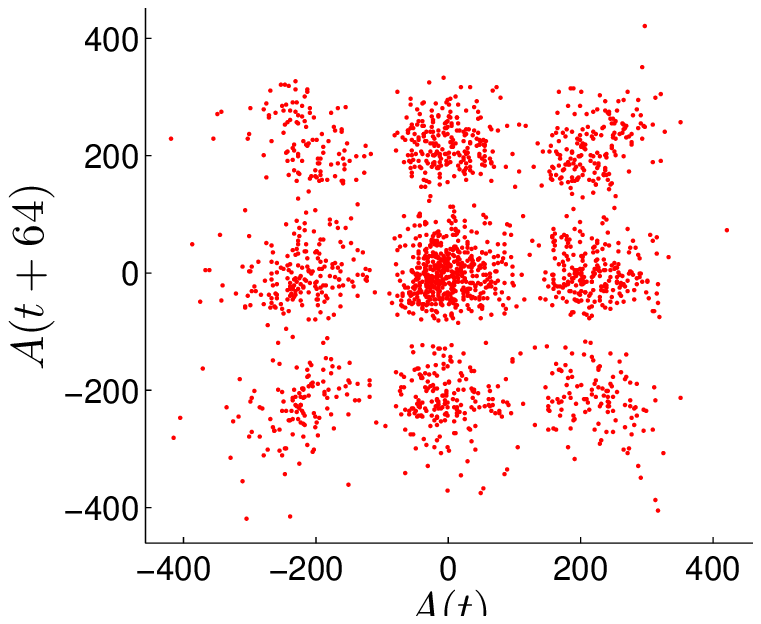}
\end{tabular}
\end{center}
\vspace*{8pt}
\caption{\label{fig33} Plots of the aggregated demand $A(t+2\cdot 2^m)$ vs. $A(t)$ for three combinations of the population size $N$ and agent memory $m$: $N=401$, $m=1$ (left), $N=1601$, $m=2$ (middle) and $N=1601$, $m=5$ (right). Simulations were done for $S=2$ and $g(x)=\sgn (x)$. Apparent preferred levels of $A(t)$ are seen as clusters of points. For $m=1$ and $m=2$ points tend to flock around diagonals indicating positive correlation for $\tau=2\cdot 2^m$.}
\end{figure}

\begin{figure}[t]
\begin{center}
\begin{tabular}{ccc}
\includegraphics[scale=.27]{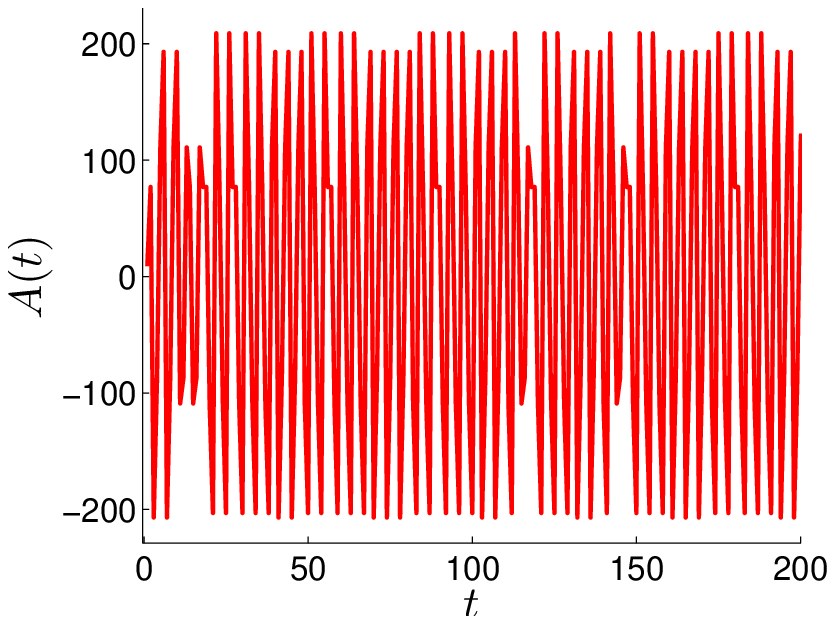} & \includegraphics[scale=.27]{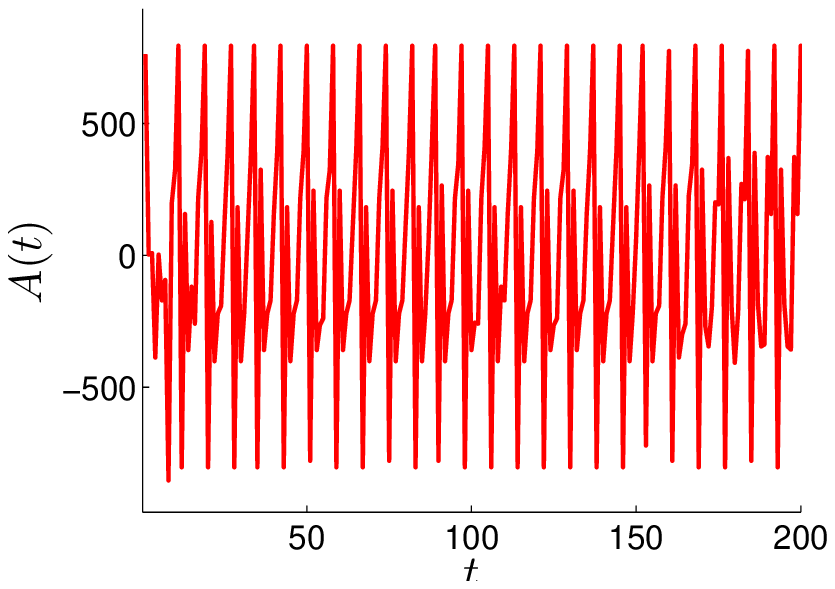} & \includegraphics[scale=.27]{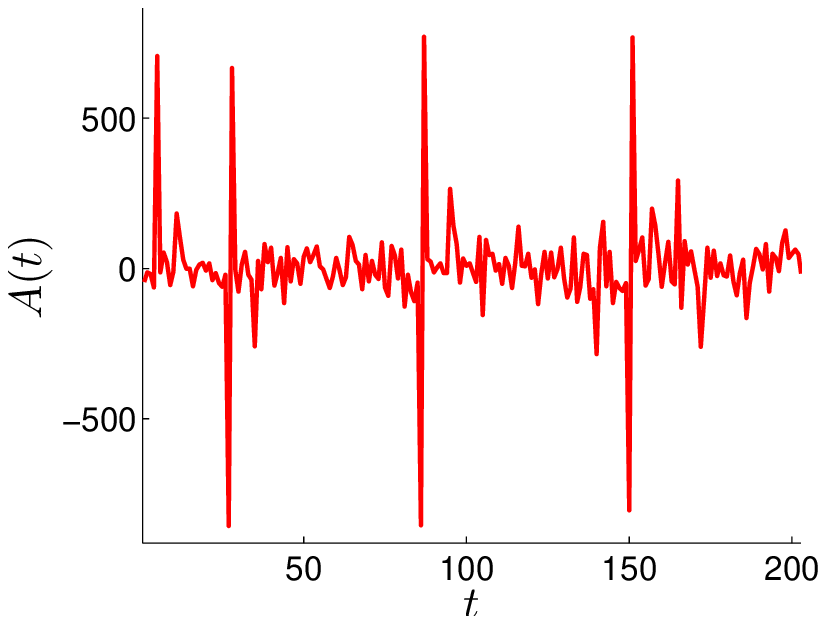}
\end{tabular}
\end{center}
\vspace*{8pt}
\caption{\label{fig34} Time evolution of the aggregated demand $A(t)$ for three combinations of the population size $N$ and agent memory $m$: $N=401$, $m=1$ (left), $N=1601$, $m=2$ (middle) and $N=1601$, $m=5$ (right). Simulations were done for $S=2$ and $g(x)=x$. Preferred values of $A$ are visible for all three games.}
\end{figure}

\begin{figure}[t]
\begin{center}
\begin{tabular}{ccc}
\includegraphics[scale=.27]{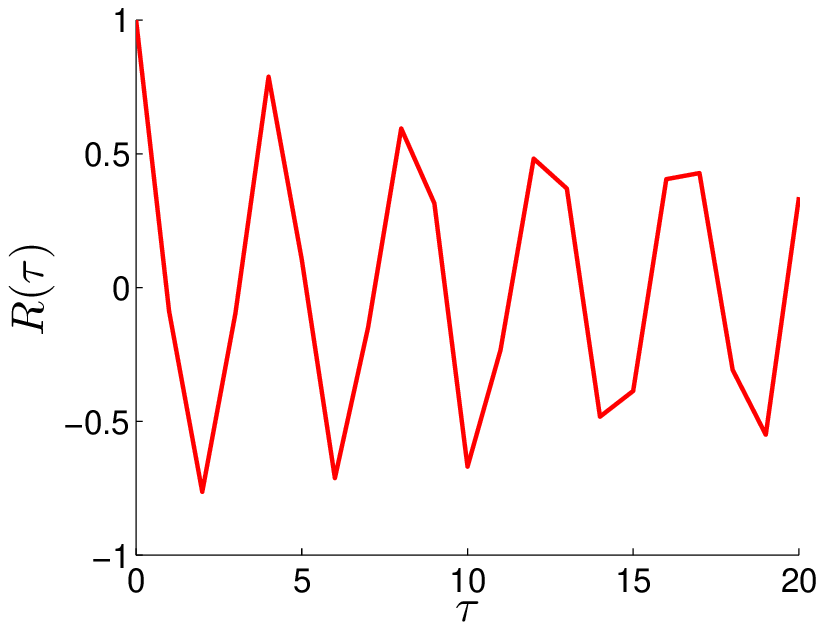} & \includegraphics[scale=.27]{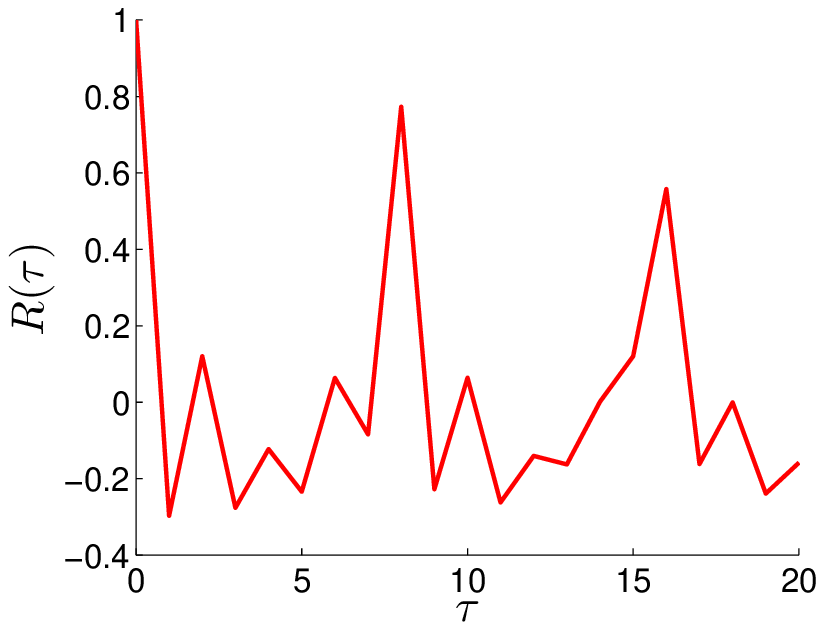} & \includegraphics[scale=.27]{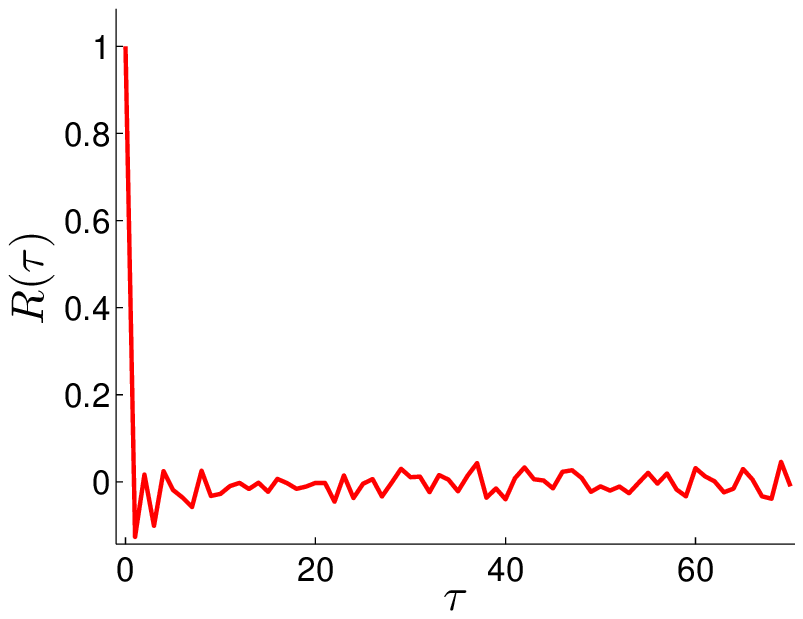}
\end{tabular}
\end{center}
\vspace*{8pt}
\caption{\label{fig35} Autocorrelation function $R(\tau)$ for three combinations of the population size $N$ and agent memory $m$: $N=401$, $m=1$ (left), $N=1601$, $m=2$ (middle) and $N=1601$, $m=5$ (right). Simulations were done for $S=2$ and $g(x)=x$. The highest values of $R$ are for $\tau = 2\cdot 2^m$, except for $\tau=0$, for all games fulfilling the $NS\gg 2^P$ condition.}
\end{figure}

\begin{figure}[t]
\begin{center}
\begin{tabular}{ccc}
\hspace{-8pt} \includegraphics[scale=.27]{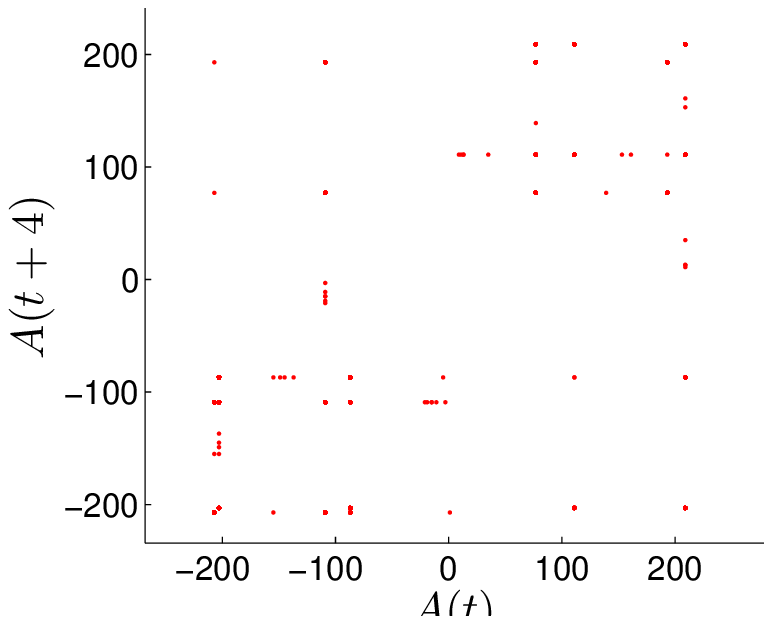} & \hspace{-8pt} \includegraphics[scale=.27]{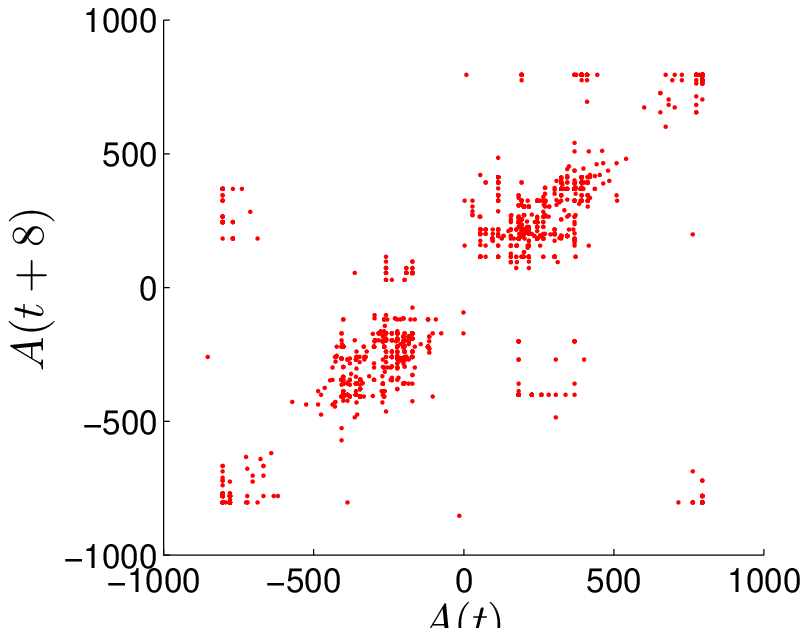} & \hspace{-8pt} \includegraphics[scale=.27]{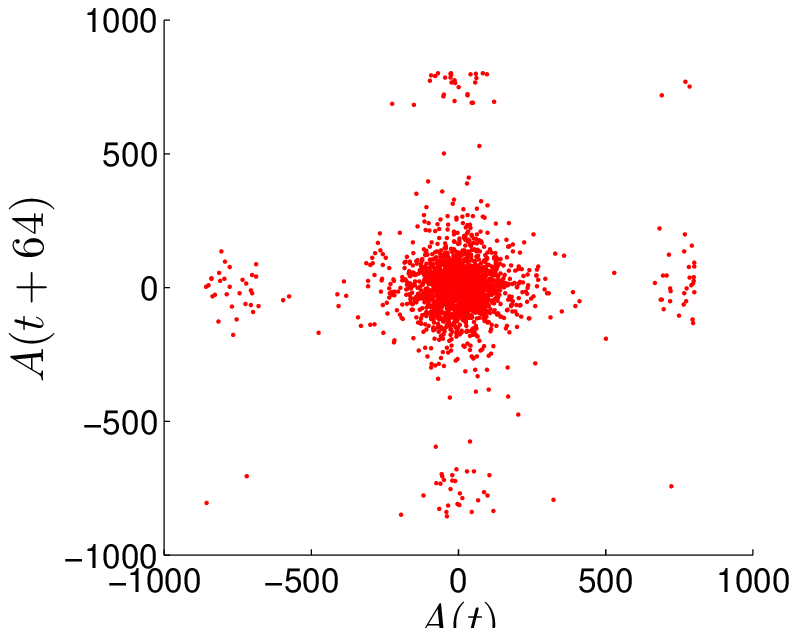}
\end{tabular}
\end{center}
\vspace*{8pt}
\caption{\label{fig36} Plots of the aggregated demand $A(t+2\cdot 2^m)$ vs. $A(t)$ for three combinations of the population size $N$ and agent memory $m$: $N=401$, $m=1$ (left), $N=1601$, $m=2$ (middle) and $N=1601$, $m=5$ (right). Simulation was done for $S=2$ and $g(x)=x$. For $m=1$ and $m=2$ points tend to flock around diagonals, indicating positive correlation, but clusterization of points is not much pronounced.}
\end{figure}

Figs \ref{fig31}, \ref{fig32} and \ref{fig33} present results for the steplike payoff function $g(x)=\sgn (x)$: the time evolution of $A(t)$, the autocorrelation function $R(\tau)$ and the scatter plots of $A(t+2\cdot 2^m)$ against $A(t)$, respectively.
The same results for the proportional payoff function $g(x)=x$ are given in Figs \ref{fig34}, \ref{fig35} and \ref{fig36}.

Even a fleeting glance at Figs \ref{fig31} and \ref{fig34} reveals regularities in $A(t)$ for both payoff functions but more regular and distinct for $g(x)=x$.
In this case their period increases with the memory length $m$ and their maximal values are equal to the half of the population size $N/2$.
This periodicity can be better seen using autocorrelation function $R(\tau)$ (cf. Figs \ref{fig32} and \ref{fig35}) where $\tau$ is the correlation time.
The autocorrelation $R$ exhibits statistically periodic peaks with periods $T=2\cdot 2^m$, as has been already observed in the efficient regime in refs \cite{zheng,jeffries_1}.
The autocorrelation is much less pronounced for games which do not meet the criterion $NS\gg 2^P$, as seen in Figs~\ref{fig32} and \ref{fig34} (right).
Relaxation of this criterion spoils periodicity of the aggregated demand.
Similar observations can be done inspecting the $A(t+2\cdot 2^m)$ {\it vs.} $A(t)$ scatter plots in Figs \ref{fig33} and \ref{fig36} where points for games fulfilling $NS\gg 2^P$ condition (left and middle panels in Figs \ref{fig33} and \ref{fig36}) are stronger flocked around diagonals. 

Another interesting feature of the aggregated demand, seen in the one-dimensional plots of $A(t)$ and better in the two-dimensional plots $A(t+2\cdot 2^m)$ {\it vs.} $A(t)$, is an existence of preferred values of $A$.
These preferred values show up as specles in the two-dimensional plots.
The specles are better focused and more numerous for $g(x)=\sgn (x)$ (Fig. \ref{fig33}) than for $g(x)=x$ (Fig. \ref{fig36}).

\begin{figure}[h]
\begin{center}
\begin{tabular}{cc}
\includegraphics[scale=.4]{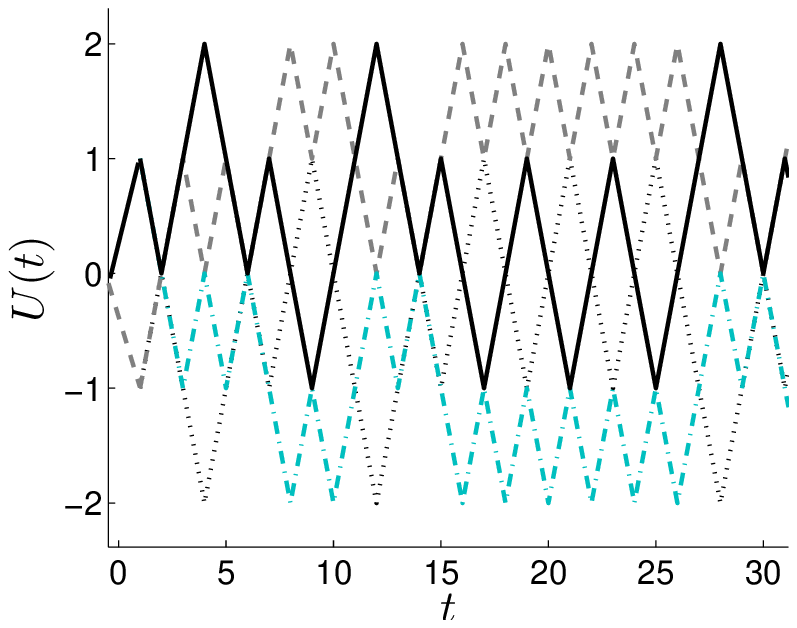} & \includegraphics[scale=.4]{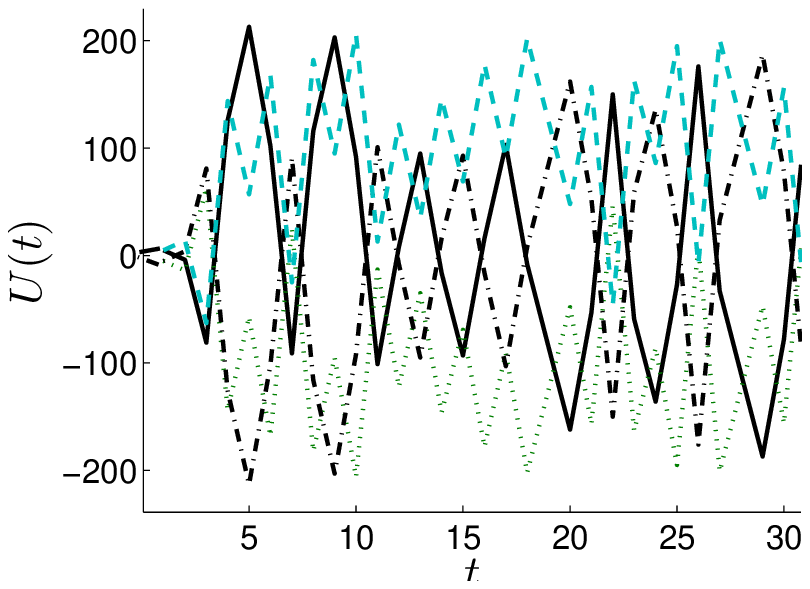}
\end{tabular}
\end{center}
\vspace*{8pt}
\caption{\label{fig41} Trajectories of the utility function $U(t)$ for all strategies of the MG with $S=2$ and $m=1$ and $N$ high enough to ensure the $NS\gg 2^P$ regime. Two payoff functions are shown: the steplike $g(x)=\sgn(x)$ (left) and the proportional $g(x)=x$ (right). Lines correspond to all different strategies. Note difference of vertical scales between panels.}
\end{figure}

Time evolution of the utility functions appears to be strongly mean-reverting processes, independently of the payoff function, as seen e.g. in Figs \ref{fig41}.
The more so, for the steplike payoff $g(x)=\sgn (x)$ the utility is bounded to rather narrow belt $-2^m\le U(t)\le 2^m$, where here and in Fig.~\ref{fig41} $U(t)$ stands for the utility for any strategy.
The formal proof of this statement is given in chapter 5.
This feature is observed for any $N$ and $S$, provided the criterion $NS\gg 2^P$ is met.

\section{The concept of state}

Since the MG represents system with many degrees of freedom, dimesionality of states is expected to be large.
In general, for each time step $t$, specification of state $x(t)$ consists of: 
\begin{itemize}
\item[A.] The history of decisions $\mu(t)$, 
\item[B.] The set of strategies of all agents $\{\alpha_n^s\}_{n=1,\ldots,N}^{s=1,\ldots,S}$,
\item[C.] The set of utilities for all strategies of all agents $\{U_{\alpha_n^s}(t)\}_{n=1,\ldots,N}^{s=1,\ldots,S}$,
\item[D.] A function relating strategies to agents: $\rho(n):n\rightarrow\Delta_n$.
\end{itemize}
Although the history of decisions $\mu(t)$ partially stores information about the past of the process, transition probabilities depend only on the present state and the process is Markovian.

Substantial reduction of the number of state parameters and simplification of state description are possible in our case.
Agents can use identical strategies \footnote{Two strategies are called different if their Hamming distance is not equal to zero. The number of pairwise different strategies is equal to $2^P$.}.
Expected number of identical strategies in the whole population behaves asymptotically, for $N\rightarrow\infty$, like $NS/2^P$.
The condition $NS\gg 2^P$ assures that the game stays in that asymptotic regime and the number of identical strategies is close to its asymptotic expected value.
Identical strategies have the same utilities over the whole game, provided the initial values of strategies are the same, e.g. $U(0)=0$, for all strategies. 
It is thus enough to take into account only reduced set of pairwise different strategies $\{\beta_i\}_{i=1}^{2^P}$ and utilities defined on them:
\begin{itemize}
\item[B.] $\{\alpha_n^s\}_{n=1,\ldots,N}^{s=1,\ldots,S}\longrightarrow \{\beta_i\}_{i=1}^{2^P}$,
\item[C.] $\{U_{\alpha_n^s}(t)\}_{n=1,\ldots,N}^{s=1,\ldots,S}\longrightarrow \{U_{\beta_i}(t)\}_{i=1}^{2^P}$.
\end{itemize} 
Concerning point D, it is sufficient to find probabilities for agents to have strategies from the set of pairwise different strategies.
The probability that given agent has any particular strategy from this set is equal to $1-(1-1/2^P)^S$.
For large $N$, the number of agents having this strategy is equal to $N(1-(1-1/2^P)^S)$.
Therefore point D, i.e. a function ascribing strategies to agents, corresponding to the agent grouping tensor $\Omega$ of ref.~\cite{jeffries_1}, can be dropped out entirely in our case.

Finally, we describe states using $\mu(t)$ and the set of utilities for the complete set of $2^P$ pairwise different strategies $\{\beta_i\}_{i=1}^{2^P}$:
\begin{eqnarray}
x(t)=[\,\mu(t),\,U_1(t),U_2(t),\ldots,U_{2^P}(t)\,].
\label{eq41}
\end{eqnarray}

Similar description of state was used in ref.~\cite{jeffries_1}.
There are, however, two important differences between their description and ours: (i) the authors of ref.~\cite{jeffries_1} introduce a functional map giving time evolution of the system in any regime, and (ii) they degenerate the game by following mean values of demand, thus making the process deterministic and Markovian, and retaining possibility to randomize it perturbatively. 
Contrary to them, we do not degenerate the game.
We consider it as a stochastic Markov process and eventually calculate the probability measure on states for the steplike payoff.

Utilities $\{U_{\beta_i}(t)\}_{i=1}^{2^P}$, considered as functions of time, are called {\it trajectories}.
In majority of cases and provided the number of observed time steps is large enough, strategies can be distinguished by their trajectories.
The sufficient condition that all $2^P$ trajectories $U_{\beta_i}(t)$ $(0\le t\le t_0)$ are distinguishable at $t_0$ is that all $2^m$ possible histories $\mu$ appear until then in a row.
On the other hand, appearance of all histories $\mu$ until $t_0$, but not necessarily exclusively, represents a necessary condition of distinguishability for trajectories.
Examples of MGs in the regime $NS\gg 2^P$ are shown in Figs~\ref{fig41} where trajectories are plotted for $m=1$ and $S=2$ and for two payoff functions further studied in this paper: $g(x)=\sgn (x)$ and $g(x)=x$.

\section{Analysis of the minority game with payoff $g(x)=\sgn (x)$}

\subsection{Finitness of the number of states}

In this chapter we demonstrate that for any $t$ the utility for any strategy is bounded from the bottom and top: $U_{min}\le U(t)\le U_{max}$, where $U_{min(max)}=-\mbox{\scriptsize (}+\mbox{\scriptsize )}2^m$.

Assume that at given time $t$ two different strategies have the same utilities.
From eqn (\ref{eq15}) for the steplike payoff function it follows that after one time step these utilities can either differ by two units or remain the same.
If the initial values of the utilities of all $SN$ strategies at $t=0$ are the same and after $\tau$ time steps at least one of them attains its extremal value, $U_{min}$ or $U_{max}$, then the trajectories cover the set of $2^m+1$ values (cf. Fig.~\ref{fig41}, left)
\begin{eqnarray}
U(\tau) & \in & \{u_l\}_{l=1}^{2^m+1} \nonumber \\
         & = & \{2^m, 2^m-2, \ldots, 2, 0, -2, \ldots, -2^m+2, -2^m\}.
\label{eq51}
\end{eqnarray}
Possible evolution scenarios leading to the values $U_{min(max)}$ can be designed by using transitions described in Appendix A.
Using this notation we have $u_1=U_{max}$ and $u_{2^m+1}=U_{min}$.
The number of different strategies characterized by the same $u_l$ is given by combinatorics as the number of trajectories starting from 0 and ending at $u_l$
\begin{eqnarray}
\#\{\beta_i: U_{\beta_i}=u_l\}=\left (\begin{array}{c} U_{max} \\ l-1 \end{array}\right ), \quad\quad i=1,\ldots,2^P.
\label{eq52}
\end{eqnarray}
The probability that the active strategy of the $n$-th agent $\alpha_n^\prime$ has utility $u_l$ is equal to
\begin{eqnarray}
{\mathcal P}\big [U_{\alpha_n^\prime}(t)=u_l\big ]=\left\{\begin{array}{lr} 1-{\mathcal P}\big [U_{\alpha_n^\prime}(t)<u_l\big ], & \quad l=1 \\ {\mathcal P}\big [U_{\alpha_n^\prime}(t)<u_{l-1}\big ]-{\mathcal P}\big [U_{\alpha_n^\prime}(t)<u_l\big ], & \quad l>1 \end{array} \right .
\label{eq53}
\end{eqnarray}
Using argumentation similar to that of ref. \cite{hart_2}, but extended to the full strategy space, one finds that
\begin{eqnarray}
{\mathcal P}\big [U_{\alpha_n^\prime}(t)<u_l\big ] & = & \prod_{s=1}^S \Big [1-{\mathcal P}\big [U_{\alpha_n^s}(t)\ge u_l\big ]\Big ] \nonumber \\
                                                 & = & \Big [ 1-\frac{\#\{\beta_i: U_{\beta_i}\ge u_l\}}{2^P}\Big ]^S,
\label{eq54}
\end{eqnarray}
where, for $t=\tau$, 
\begin{eqnarray}
\#\{\beta_i: U_{\beta_i}\ge u_l\}=\sum_{j\ge l} \left (\begin{array}{c} U_{max} \\ j-1 \end{array}\right ).
\label{eq55}
\end{eqnarray}
Denoting ${\mathcal P}_{max(min)}={\mathcal P}\big [U_{\alpha_n^\prime}(\tau)=U_{max(min)}\big ]$, one sees from eqn (\ref{eq53}) that ${\mathcal P}_{max}>{\mathcal P}_{min}$.
We notice that for any utility $u_l$, different than $U_{min}$ or $U_{max}$, the number of different strategies (\ref{eq52}) is even.
Even more, a half of strategies corresponding to each level $U_{min}< u_l < U_{max}$ suggest the opposite action than another half.
According to eqn (\ref{eq53}), if two (or more) strategies have the same utility, then all have the same probability to be the best strategies for the $n$-th agent.
This means that, if one excludes the best and the worst strategies, a half of remaining strategies recommends the same action as the best or the worst strategy.
Hence the probability that an agent plays according to the strategy suggesting the same action as the best strategy is equal to
\begin{eqnarray}
{\mathcal P}\big [a_{\alpha_n^\prime}(\tau)=a_{\alpha^B}(\tau)\big ] & = & {\mathcal P}_{max}+\frac{1}{2}\big (1-{\mathcal P}_{max}-{\mathcal P}_{min}\big ) \nonumber \\
 & = & \frac{1}{2}\big (1+{\mathcal P}_{max}-{\mathcal P}_{min}\big ),
\label{eq56}
\end{eqnarray}
where $\alpha^B(t)$ is the best strategy from the whole set of strategies in the game, i.e. $U_{\alpha^B(t)}=u_1$, and $1-{\mathcal P}_{max}-{\mathcal P}_{min}$ refers to the probability that the agent's best strategy is neither the worst nor the best of all strategies.
The factor $\frac{1}{2}$ reflects that a half of strategies with non-extremal utilities suggest the same action as the best one.
As ${\mathcal P}_{max}>{\mathcal P}_{min}$, from eqn (\ref{eq56}) it follows that if one of strategies has the utility $U_{max}$, then more than half of the population plays according to the best strategy.
Subsequently, this subpopulation loose and gets the negative payoff.
The rest are the winners and get the positive payoff.
This mechanism bounds the utility to stay between $U_{min}$ and $U_{max}$.
In addition, we know the formula for the fraction of agents playing the same action.

\subsection{Representation of the minority game as the Markov process}

\subsubsection{Case $m=1$}

\begin{figure}[h]
\begin{center}
\includegraphics[scale=.5]{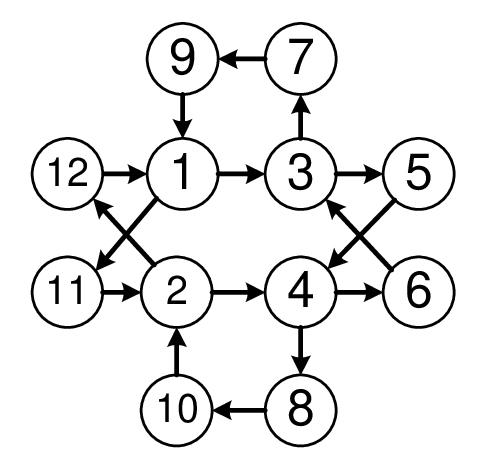}
\end{center}
\vspace*{8pt}
\caption{\label{fig52} Diagram of the Markov chain representation of the MG in the efficient regime for $m=1$. If transitions to two states are possible from a given state, both {\it a priori} transition probabilities are the same. This happens for $x_1\rightarrow x_{3,11}$, $x_2\rightarrow x_{4,12}$, $x_3\rightarrow x_{5,7}$ and $x_4\rightarrow x_{6,8}$.}
\end{figure}

In this case the complete specification of states and calculation of the transition matrix are relatively easy.
All strategies are listed in Tab. \ref{tab1}
\begin{table}[h]
\begin{center}
\begin{tabular}{|r|rrrr|} \hline
$\mu$ & $\alpha_1$ & $\alpha_2$ & $\alpha_3$ & $\alpha_4$ \\ \cline{1-5} \hline
 -1 & -1 & -1 &  1 & 1 \\
  1 & -1 &  1 & -1 & 1 \\ \hline
\end{tabular}
\end{center}
\caption{\label{tab1}Strategies for $m=1$}
\end{table}
and states are listed in Tab. \ref{tab2}.
\renewcommand{\arraystretch}{1.3}
\begin{table}[h]
\begin{center}
\begin{tabular}{|r|rrrrr|r|r|} \hline
 & $\mu$ & $U_1$ & $U_2$ & $U_3$ & $U_4$ & ${\mathcal P}(x_i)$ & ${\mathbb E}\,A(x_i)$ \\ \cline{1-8} \hline
 $x_1$ &    -1 &  0 &  0 &  0 &  0 & $\frac{1}{8}$ & 0 \\ 
 $x_2$ &     1 &  0 &  0 &  0 &  0 & $\frac{1}{8}$ & 0 \\
 $x_3$ &     1 & -1 & -1 &  1 &  1 & $\frac{1}{8}$ & 0 \\
 $x_4$ &    -1 &  1 & -1 &  1 & -1 & $\frac{1}{8}$ & 0 \\
 $x_5$ &    -1 &  0 & -2 &  2 &  0 & $\frac{1}{16}$ & $\frac{3}{8}N$ \\
 $x_6$ &     1 &  0 & -2 &  2 &  0 & $\frac{1}{16}$ & $-\frac{3}{8}N$ \\
 $x_7$ &     1 & -2 &  0 &  0 &  2 & $\frac{1}{16}$ & $\frac{3}{8}N$ \\
 $x_8$ &    -1 &  2 &  0 &  0 & -2 & $\frac{1}{16}$ & $-\frac{3}{8}N$ \\
 $x_9$ &    -1 & -1 & -1 &  1 &  1 & $\frac{1}{16}$ & $\frac{1}{2}N$ \\
 $x_{10}$ &  1 &  1 & -1 &  1 & -1 & $\frac{1}{16}$ & $\frac{1}{2}N$ \\
 $x_{11}$ & -1 &  1 &  1 & -1 & -1 & $\frac{1}{16}$ & $-\frac{1}{2}N$ \\
 $x_{12}$ &  1 & -1 &  1 & -1 &  1 & $\frac{1}{16}$ & $-\frac{1}{2}N$ \\ \hline
\end{tabular}
\end{center}
\caption{\label{tab2}States $x_i$ $(i=1,\ldots,12)$, their probabilities ${\mathcal P}(x_i)$ and demands for $m=1$. The ${\mathbb E}\,A(x_i)$ stands for the expected value of $A$ for the state $x_i$. The ${\mathcal P}$ and ${\mathbb E}$ represent {\it a priori} values, i.e. before strategies are assigned to agents. After game initialization these values may become different and depend on realization of the game but the sequence of states is preserved.}
\end{table}
At the beginning of the game we assume no {\it a priori} knowledge, so that all utilities are equal to zero, and two initial states are possible: $x_1$ and $x_2$.
For these two states the values of $\mu(t)$ are different.
Subsequent time evolutions depend on ratios between numbers of agents playing $+1$ or $-1$ actions and are illustrated in Figs \ref{fig51} and described in detail in Appendix A. 
These states and transitions are sufficient to define a memoryless representation of the MG with a transition graph displayed in Fig. \ref{fig52}.
Some of its states have the same expected demand ${\mathbb E}\,A$ over realizations of the game, e.g. ${\mathbb E}\,A(x_i)=0$ ($i=1,...,4$), as the same numbers of agents play according to strategies recommending opposite actions.
Using formulas (\ref{eq53}-\ref{eq55}) we can find ${\mathbb E}\,A$ for all states (cf. Tab. 2), consistently with observations in Fig. \ref{fig33}, where five clusters on the diagonal are found around values from Tab.~\ref{tab2}.

Our process is a stationary Markov chain for which the stationary Master Equation can be solved with respect to the state probabilities.
Their values are given in Tab.~\ref{tab2}, in the column marked ${\mathcal P}(x_i)$ $(i=1,\ldots,12)$.
The state probabilities from Tab.~\ref{tab2} can be also used to find statistical periods of the demand
\begin{eqnarray}
{\mathcal P}\big[A(t)=A(t+\tau)\big] & = & \sum_{ij}\,\delta\big[A\big(x_j(t+\tau)\big),A\big(x_i(t)\big)\big] \nonumber \\
& \cdot & {\mathcal P}\big[x_j(t+\tau)\,|\,x_i(t)\big]\cdot {\mathcal P}\big[x_i(t)\big],
\label{eq61}
\end{eqnarray}
where
\begin{eqnarray}
\delta(x,y)=\left\{\begin{array}{lc} 1, & x=y \\ 0, & \mbox{\small otherwise} \end{array} \right . 
\end{eqnarray}
The maximal value of $7/16$ is found for $\tau=4$ and this explains why the largest correlation is found also for $\tau =4$.

\subsubsection{Case $m>1$}
Any MG with $m>1$ in the efficient regime can be represented as a Markov process with a finite number of states.
The same method as for $m=1$, but more demanding computationally, can be used to calculate state probabilities. 

\section{Analysis of the minority game with payoff $g(x)=x$}

Contrary to the MG with the steplike payoff $g(x)=\sgn (x)$, in case of the proportional payoff $g(x)=x$ the pairwise different strategies with identical utilities are unlikely (cf. Fig. \ref{fig41}).
This means that the probabilities that the pairwise different strategies have the same utility is small compared to the case of $g(x)=\sgn (x)$.
Consequently, the probability that an agent has a freedom of choice of the next state is negligible for $g(x)=x$.
This means that such game is in a sense less stochastic than for $g(x)=\sgn (x)$.
Nevertheless, the game is still periodic because the number of states is finite.
A persuasive explanation of periodicity is proposed by the authors of ref. \cite{jeffries_1} using de Bruijn representation of the memory sequences $\mu$.
Here we extend their analysis and explain peaks of $A(t)$ and their frequency using two approaches based on the utility analysis.

\subsection{First approach}

\begin{figure}[h]
\begin{center}
\begin{tabular}{cc}
\includegraphics[scale=.35]{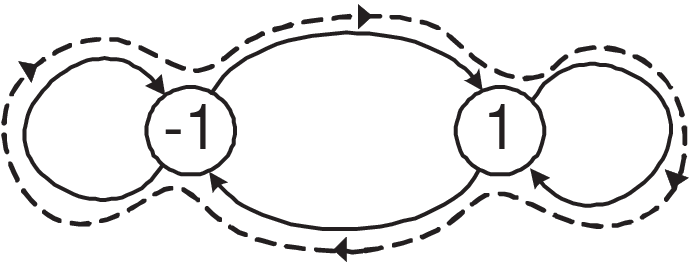} & \hspace{1cm} \includegraphics[scale=.35]{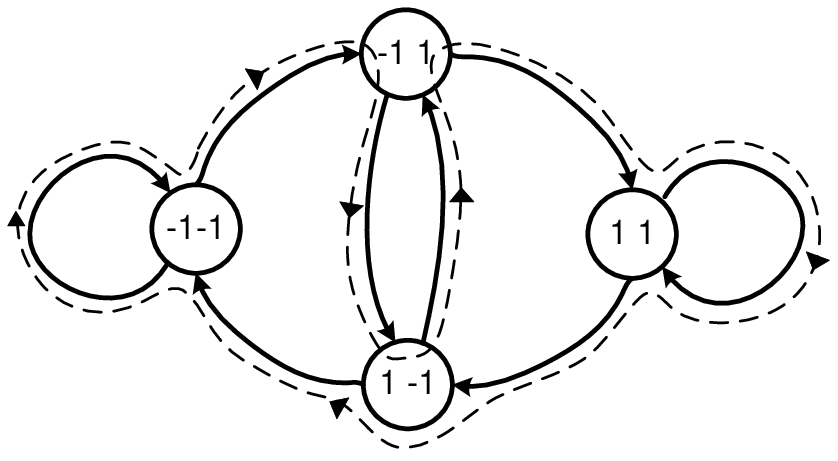}
\end{tabular}
\end{center}
\vspace*{8pt}
\caption{\label{fig61} De Bruijn graphs of orders $m=1$ (left) and $m=2$ (right). Dashed lines represent examples of the Euler trails on the graph: one trail for $m=1$ (left) and one of two possible Euler trails for $m=2$ (right)}
\end{figure}

Dynamics of the MG can be efficiently studied using de Bruijn graphs, as shown in ref. \cite{challet_3}.
The decision history $\mu(t)$ is a sequence of $m$ minority actions
\begin{eqnarray}
\mu(t)=\big[a^\ast(t-m),a^\ast(t-m+1),\ldots,a^\ast(t-1)\big].
\label{eq62}
\end{eqnarray}
The $\mu(t+1)$ is obtained by adding $a^\ast(t)$ to the right and deleting $a^\ast(t-m)$ from the left of the vector (\ref{eq62}), such that there are two possible successors $\mu(t+1)$ of $\mu(t)$.
If one history can be obtained from another one using this procedure, then the latter has a directed edge to the former one.
Histories may be represented by labeled edges.
These rules define de Bruijn graph of the order $m$.
Examples for $m=1$ and $m=2$ are given in Figs~\ref{fig61}.

Histories in MGs are not equiprobable \cite{challet_3}.
Among all paths on the de Bruijn graph of the game, Euler paths define the shortest sequence of histories where each strategy looses and wins equally likely.
In the non-Eulerian paths some histories are more frequent and therefore some strategies are more profitable.
We show in the following that in the efficient mode the non-Eulerian paths are rare compared to the Eulerian ones.

For the proportional payoff, prevalent number of strategies have unique utility.
In such a case, the probability (\ref{eq53}) for the active startegy $\alpha_n^\prime$ can be simplified (cf. also ref. \cite{hart_2})
\begin{eqnarray}
{\mathcal P}\big [U_{\alpha_n^\prime}(t)=u_l\big ]= \Big(1-\frac{l-1}{2^{P}}\Big)^S-\Big(1-\frac{l}{2^{P}}\Big)^S, \quad\quad l\ge 1.
\label{eq63}
\end{eqnarray}
Consider the case when $A$ is the largest possible.
Since $\{u_l\}$ is a sorted list of utilities, this is possible if the first $l/2$ strategies in this list suggest actions opposite to the last $l/2$.
Then the probability of an action suggested by the best strategy is equal to
\begin{eqnarray}
{\mathcal P}\big[a_{\alpha_n^\prime(t)}=a_{\alpha^B(t)}\big] & = & \sum_{l=1}^{2^{P-1}} {\mathcal P}\big[U_{\alpha_n^\prime(t)}=u_l\big] \nonumber \\
                                                   & = & 1-\frac{1}{2^{S}}.
\label{eq64}
\end{eqnarray}
This means that for large $NS$ for about $N(1-\frac{1}{2^{S}})$ agents their active strategy is the same as the best strategy and the expected absolute value of the aggregated demand is equal to
\begin{eqnarray}
|A|=N\Big(1-\frac{1}{2^{S-1}}\Big).
\label{eq65}
\end{eqnarray}
In particular, if $S=2$ then $|A|=N/2$.

There is also more fundamental reason that the order of strategies in the list appears such that two halves of the list suggest opposite actions.
We noticed that large fluctuation of $A$ is only possible if the game is in one of two de Bruijn nodes called {\it homogeneous}, i.e. consisting of identical symbols: $\mu_{h1(2)}=\big[-(+)1,\ldots,-(+)1\big]$. 
Interesting enough, peaks are observable only after one of the homogenous histories, but not after both, as explained technically in Appendix B.

Since high $A(t)$ appears only after the history $\mu_C$, we have just two transitions in the Eulerian path that starts from this history.
From this it follows that the frequency of peaks is equal to
\begin{eqnarray}
f & = & \frac{2}{2^{m+1}} \nonumber \\
  & = & \frac{1}{2^m},
\label{eq68}
\end{eqnarray}
in agreement with our simulations.
The value $2^{m+1}$ is the length of the Euler path and it corresponds to the period of $A$ observed in Fig.~\ref{fig36}.

Our argumentation becomes strict and eqn~(\ref{eq64}) is exact in the efficient mode when $NS\gg 2^P$, ideally in the limit $NS\rightarrow\infty$.
But we also observe cyclic peaks of demand for $N=1601$ and $m=5$, when the efficiency condition is not met (cf. Fig.~\ref{fig34}, right).
In fact, the condition $NS\gg 2^P$ can be slacken off to the requirement that the population is numerous enough that the game is in the herd mode. 
Games in that mode do not follow Eulerian paths because for smaller $N$ the pool of strategies is too sparse and some histories occur more frequently.
Nevertheless, the mechanism of peak creation is approximately preserved, as long as $N$ is large enough to cause the split of utilities into two groups. 

\subsection{Second approach}

\begin{figure}[h]
\begin{center}
\begin{tabular}{cc}
\includegraphics[scale=.39]{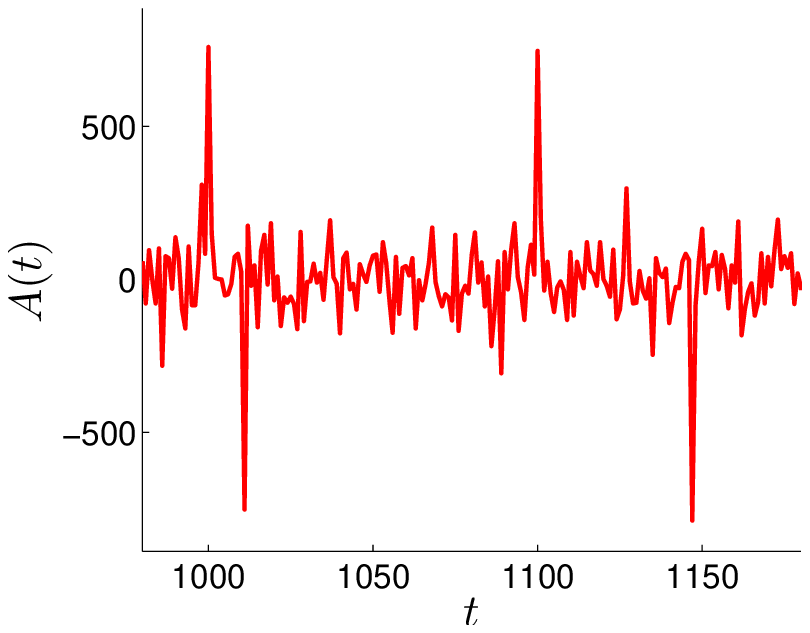} & \includegraphics[scale=.39]{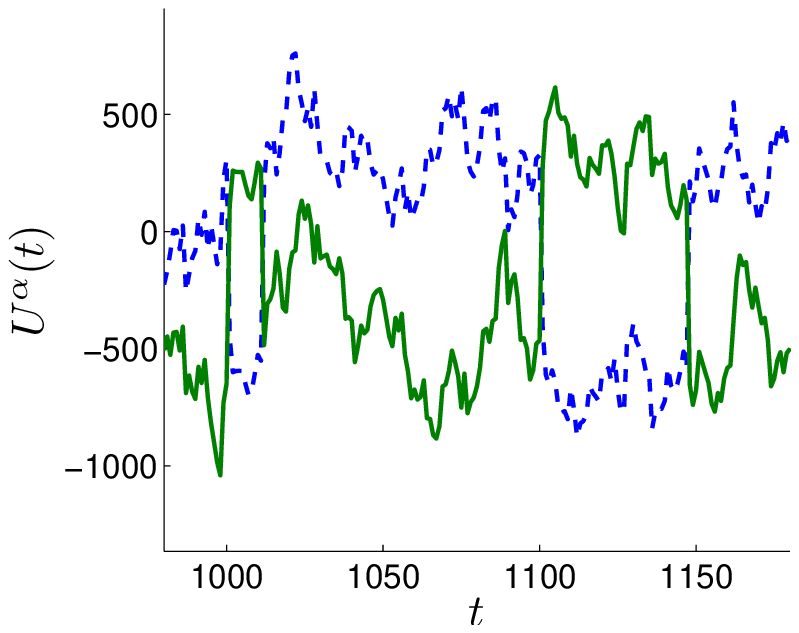} \\
\includegraphics[scale=.39]{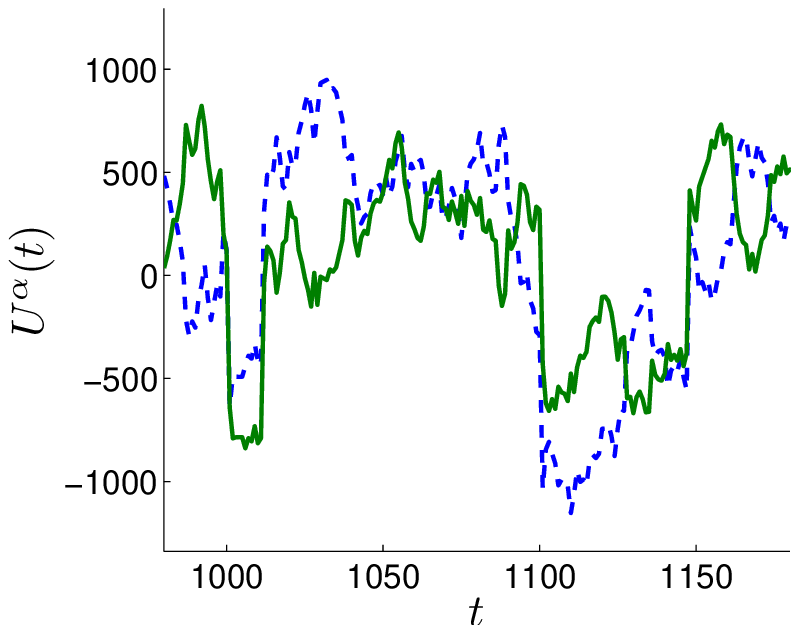} & \includegraphics[scale=.39]{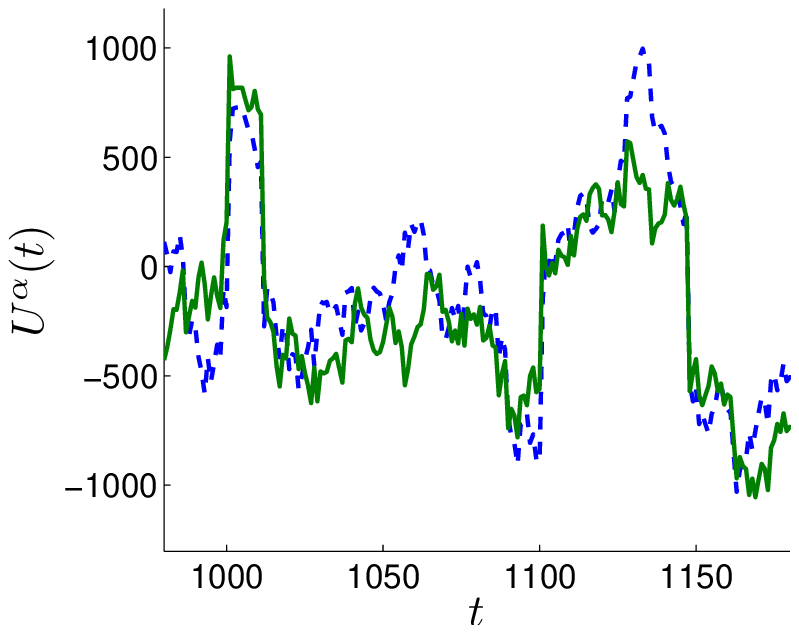}
\end{tabular}
\end{center}
\vspace*{8pt}
\caption{\label{fig64}The time evolution of the aggregated demand (upper left) and utilities for three cases: an agent with one high- and one low-utility strategy (upper right), two high-utility strategies (lower left) and two low-utility strategies (lower right) at $t=1000$. These three cases may be quantitatively distinguished using the values of utilities at $t=1000$, corresponding to the location of the first maximum of $A(t)$ in the upper left panel. Simulation was performed for the MG with $N=1601$, $S=2$, $m=5$ and $g(x)=x$.}
\end{figure}
At any time a somewhat simpler explanation may be given by dividing strategies into two categories: the {\it good} with the positive payoff, and {\it bad} with negative \cite{wawrzyniak}.
Probability that an agent has no good strategies, or at least one good, is equal to $\frac{1}{2^{S}}$ and $1-\frac{1}{2^{S}}$, respectively.
Rapid fluctuations of demand are transferred to similar fluctuations of the utility.
The $A(t_1)$ fluctuates after the history $\mu_C=\mu(t_1)$ when the strategies with higher utility indicate identical actions.
If $A(t_1)$ strongly fluctuates, then at $t_1+1$ about $N(1-\frac{1}{2^{S}})$ agents have at least one strategy with high utility and they choose it.
Strategies split into two groups of high and low utility with a gap between these two groups (cf. Fig.~\ref{fig62} in Appendix B).
Strategies with high/low utility do not suggest the same actions, provided $\mu\ne\mu_C$, and therefore no peak of $A$ is generated.
The $\mu_C$ has a non-vanishing probability to reappear at some $t_2>t_1$.
All agents belonging to the group with at least one high-utility strategy tend to react identically and $A(t_2)$ fluctuates maximally, i.e. $A(t_2)=N(1-\frac{1}{2^{S-1}})$.
This is illustrated in Fig.~\ref{fig64} (upper left), where for $S=2$ we have $A(t=1000)=\frac{N}{2}$.
At $t_2$, all strategies with high $U(t_2)$ fail and get the penalty $-A(t_2)$, whereas those with low $U(t_2)$ are rewarded with  $A(t_2)$.
After $t_1$ agents are divided into three groups, provided $S=2$: the group with two good strategies, with one good and one bad, and with two bad.
As seen in Fig.~\ref{fig64}, at $t=1000$ a quarter of the population with two high-utility strategies evolves into two low-utility group (lower left), and {\it vice versa} for another quarter with two initially low-utility strategies (lower right).
Remaining half of the population just swaps utilities of their strategies (upper right).

Results showing periodicity of $A(t)$ from simulations become closer to the theoretical results for large $NS/2^P$ ratio.
If it is small, then the game hardly follows the Eulerian path and peaks of $A(t)$ appear randomly. 

\section{Stochasticity of the game depends on initial conditions}

We assumed that $U_{\alpha_n^s}(t=0)=0$ for all $\alpha_n^s$.
This assumption seems natural as reflecting no {\it a priori} preference for any strategy.
However, it appears to be critical for the MG dynamics for $g(x)=\sgn (x)$.
Stochastic transitions mentioned in chapter 5.2 show up for the degenerate state, i.e. more than one strategy with the same utility.
Removing this ambiguity suppresses stochasticity and the game becomes deterministic.
In such a case, our simplified description of the state fails because strategies have unique utilities and cannot be aggregated.
Consequently, the Markovian treatment is no longer useful but its description in terms of de Bruijn graphs becomes interesting.
In particular, the game follows the Eulerian path on de Bruijn graph.
In case of the proportional payoff $g(x)=x$, the game is just deterministic and follows one of the Eulerian paths.

\section{Conclusions}

We studied the MG in the efficient mode.
We observe interesting collectivity in agent behaviour in this mode.
Depending on the payoff function $g(x)$, the game is driven by different dynamics which requires different methods of the analysis.
In case $g(x)=\sgn (x)$, provided the population $N$ is large enough to assure $NS\gg 2^P$, the MG can be described in terms of the Markov process with the finite number of states, where transitions may be both stochastic and deterministic.
This representation completely defines dynamics of the game in the stationary regime and allows for the calculation of state occupancies and other observables.
The Markov representation provides with an explanation of the periodicity and preferred levels of the aggregate demand $A(t)$.
In practical terms this approach is tough for $m>1$ due to the large number of states.
We failed to find any relation between the memory length $m$ and total number of states.
Neither the simplified concept of state nor the Markov process description are valid if the initial preference is given to any strategy.

For the proportional payoff $g(x)=x$, stochasticity of transitions disappears but one still observes periodicity. 
One also observes distinct peaks of the aggregated demand, exhibiting height equal to a half of the population, assuming $S=2$. 
In the herd regime, there always exists a history $\mu_C$ for which $1-\frac{1}{2^{S}}$ of agents react identically and this is seen in the peak $A(t)=N(1-\frac{1}{2^{S-1}})$. 
We provided with two compatible explanations of these phenomena.
The first uses the ordered list $\{ u_l\}$  of $2^P$ strategies and is similar to the reasoning for $g(x)=\sgn (x)$.
The second approach is a simplification of the first one to the case when only two classes of strategies are used instead of all $2^P$ classes.
The second approach was also successfully exploited in our analysis of the multi-market minority game \cite{wawrzyniak}.

We studied games with full strategy space.
Some authors, e.g. refs \cite{challet_2,li}, reduce strategy space and reproduce many features of the full MG, e.g. behaviour of $\sigma(A)^2/N$.
This trick, however, has serious drawbacks since it reduces the number of states in the Markov description of the game and significantly affects its time evolution.
For $g(x)=\sgn (x)$, the Markov representation is oversimplified by such reduction.

It this work we focused on theoretical issues of the MG with real histories.
We did not elaborate on application of our model to real-life systems, as e.g. financial markets.
At the moment, applications are more discussed by other authors \cite{challet_5,challet_6,jeffries_2,tadeschi}.
Perhaps the most general mathematical description of MGs with real histories is given in ref. \cite{coolen} using the generating functional approach.

\section*{Acknowledgments}

Results presented in this paper were obtained using computational grid build
in the framework of the project INFO-RI-222667 {\it Enabling grids for
E-science} funded by the European Commission in the 7th Framework Program.

\vspace*{5pt}

\appendix

\section{Transition scenarios for $\mathbf{m=1}$ minority game with $\mathbf{g(x)=\sgn (x)}$}

\vspace{5pt}

\begin{figure}[h]
\begin{center}
\includegraphics[scale=.35]{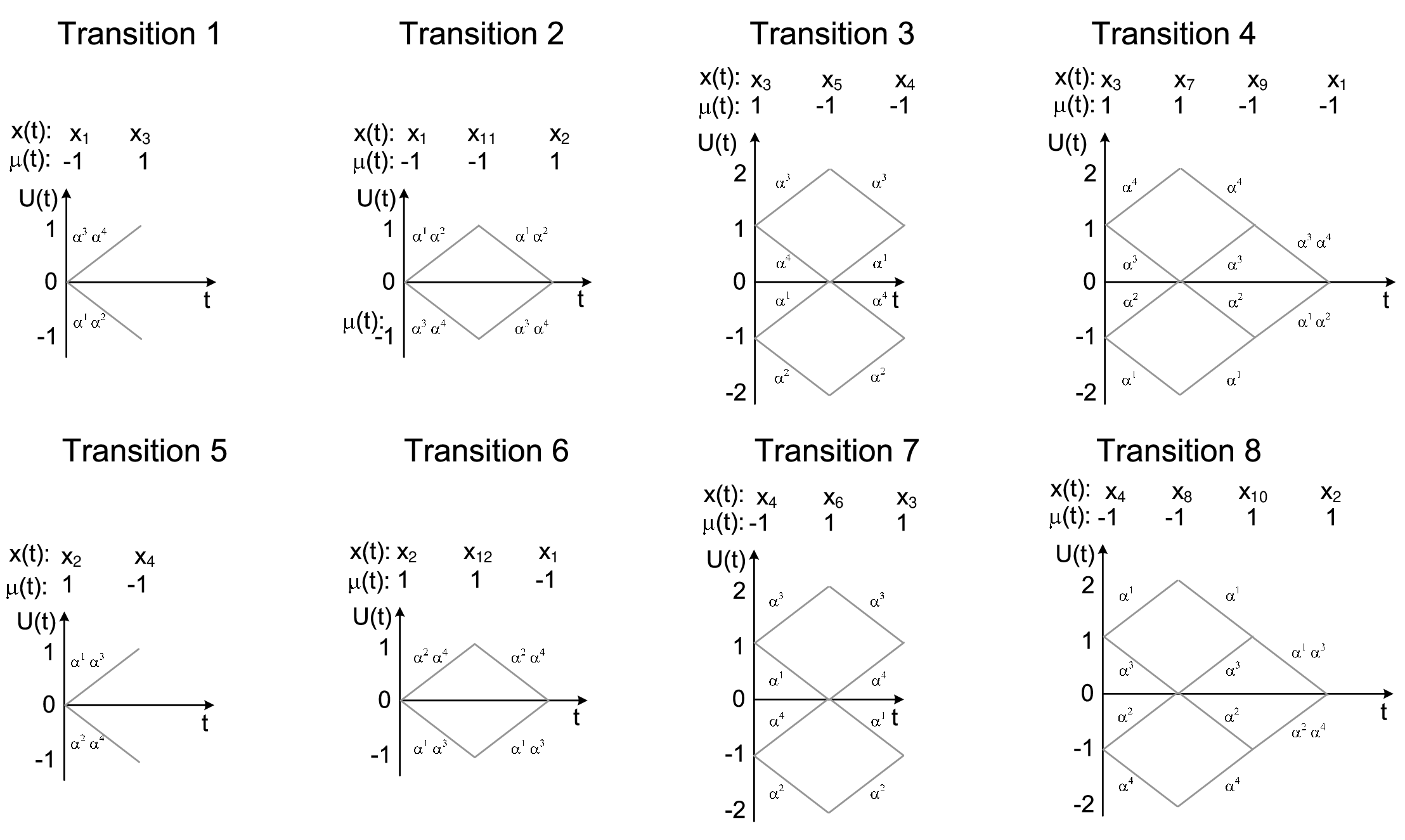} 
\end{center}
\vspace*{8pt}
\caption{\label{fig51} Trajectories of utilities for $m=1$.}
\end{figure}
Possible transition scenarios for the $m=1$ MG, represented as the Markov chain, are illustrated in Figs \ref{fig51}.
At the beginning of the game all utilities are equal to zero.
Depending on the history $\mu$, only two initial states can exist: $x_1=[-1,0,0,0]$ and $x_2=[1,0,0,0]$.
For each of these two states two further scenarios are equally possible, because the utilities of corresponding strategies are the same.
The choice depends on the ratio between numbers of agents in two groups: one with $a=1$ and another one with $a=-1$.
These scenarios are as follows.
\begin{itemize}
\item[\underline{Transition 1}]
~\\
Being in the state $x_1$, the majority of agents use strategies suggesting $a=-1$.
Then

\begin{itemize}
\item the minority action in the next step is $a^\ast=1$,
\item strategies $\alpha_1$ or $\alpha_2$ give negative payoff,
\item strategies $\alpha_3$ and $\alpha_4$ give positive payoff.
\end{itemize}
The system goes to the state $x_3=[1,-1,-1,1,1]$ (cf. Fig.~\ref{fig51}, Transition~1) where $U_{\alpha_3}=U_{\alpha_4}=1$ and these strategies suggest different actions on the last history $\mu=1$.
Similarly, there are two strategies with the utilities $U_{\alpha_1}=U_{\alpha_2}=-1$ suggesting different actions on $\mu=1$.
Hence, there are two equiprobable scenarios, further described as Transitions 3 and 4.
\item[\underline{Transition 2}]
~\\
Being in the state $x_1$, the majority of agents use strategies suggesting $a=1$.
Then
\begin{itemize}
\item the minority action in the next step is $a^\ast=-1$,
\item strategies $\alpha_3$ or $\alpha_4$ give negative payoff,
\item strategies $\alpha_1$ and $\alpha_2$ give positive payoff.
\end{itemize}
The system goes to the state $x_{11}=[-1,1,1,-1,-1]$ (cf. Fig.~\ref{fig51}, Transition~2) where $U_{\alpha_1}=U_{\alpha_2}=1$ and give the same actions on the last history $\mu=-1$.
Most of agents use these strategies (e.g. $3/4$ of the population, provided $S=2$) and the sole possibility is that the system goes to the state $x_2$.
\item[\underline{Transition 3}]
~\\
Being in the state $x_3$, the majority of agents use strategies suggesting $a=1$ and the system passes to $x_5$.
In this state $U_{\alpha_3}=U_{max}$ and $U_{\alpha_2}=U_{min}$ (cf. Fig.~\ref{fig51}, Transition~3).
According to the reasoning from section 5.1, if one utility attains its maximal or minimal value, most agents use strategies suggesting the same action as the best strategy.
Consequently, there is only one scenario possible in $x_5$: the best strategy, and all strategies giving the same output as the best one, loose and the system goes to the state $x_4$.
\item[\underline{Transition 4}]
~\\
Another possibility in $x_3$ is that most of agents decide $a=-1$ and the system goes to $x_7$.
In this state $U_{\alpha_4}=U_{max}$ and $U_{\alpha_1}=U_{min}$ (cf. Fig.~\ref{fig51}, Transition~4).
Subsequently, the best strategy, and all strategies giving the same output as the best one, loose and the system goes to the state $x_9$.
In $x_9$ both best strategies suggest the same for the last history $\mu=-1$.
The majority of the population uses one of these best strategies and the system moves to $x_1$.
\item[\underline{Transition 5--8}]
~\\
These transitions are analogical to Transitions 1--4, but the initial state is $x_2$.
\end{itemize}

\vspace{7pt}

\section{Algorithm generating strong demand fluctuations}

\vspace{5pt}

\begin{figure}[t]
\begin{center}
\includegraphics[scale=.47]{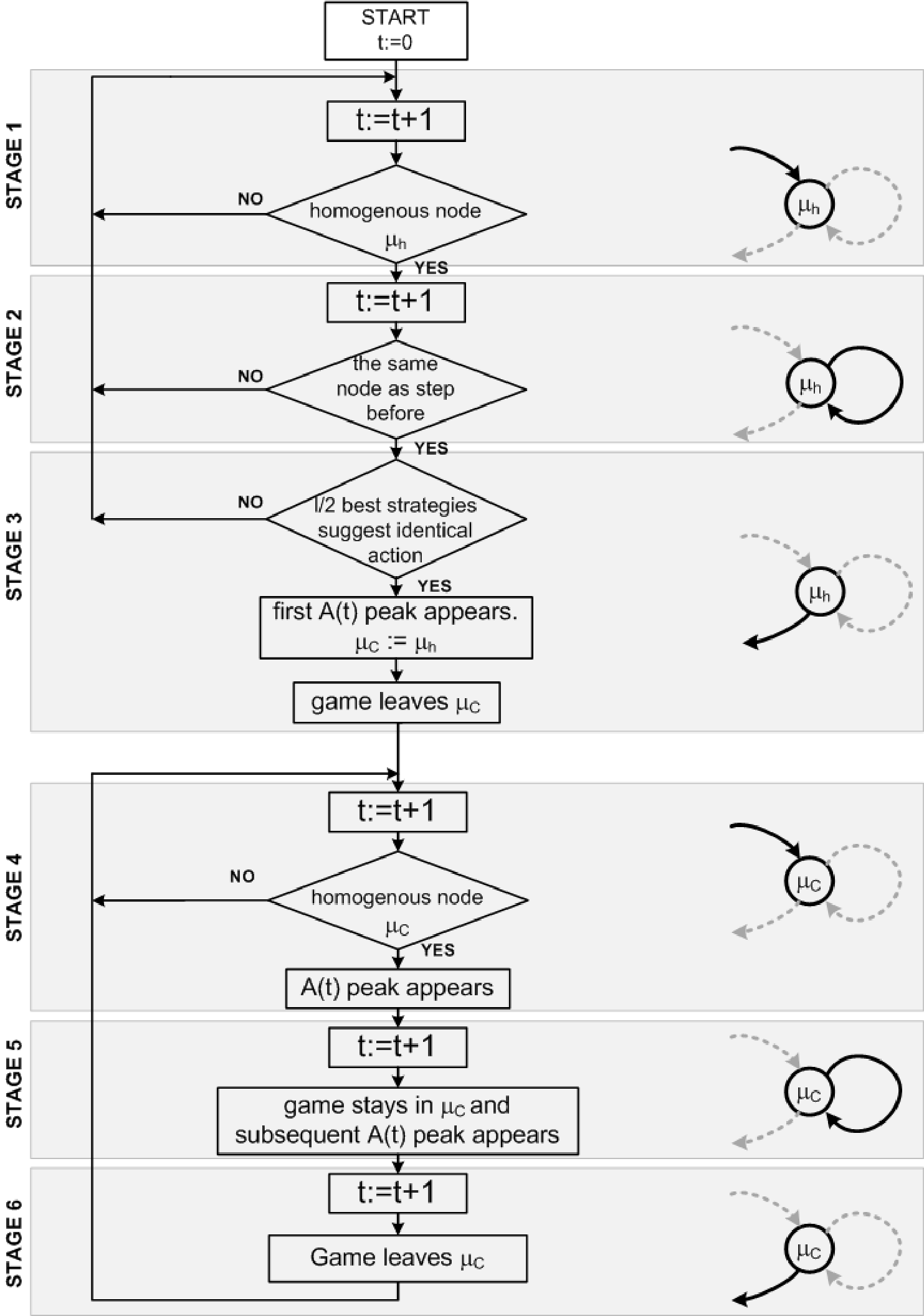}
\end{center}
\vspace*{8pt}
\caption{\label{figa1} The flow chart of the MG evolution algorithm, illustrating appearance of distinct peaks of demand.}
\end{figure}

In Fig.~\ref{figa1} we present the flow chart illustrating appearance of strong fluctuations of $A(t)$.
Below we describe the algorithm step by step.
First three stages lead to the first peak. 
Next steps explain why the subsequent peaks follow each other and why they have opposite signs.
\begin{itemize}
\item[\underline{Stage 1}]
~\\
If $A(t_1)$ stands for the first peak of demand then three prior conditions have to be fulfilled.
The first is that $\mu(t_1 - 1)=\mu_{h1(2)}$, where $\mu_{h1(2)}=[-(+)1,\ldots,-(+)1]$ is a homogeneous node.
\item[\underline{Stage 2}]
~\\
It is also required that at $t_1-1$ majority of agents decides to change the node.
If this is fulfilled then the minority action is
\begin{eqnarray}
a^\ast(t_1-1)=\left\{ \begin{array}{rr} -1, & \quad \mu(t_1-1)=\mu_{h1} \\
1, & \quad \mu(t_1-1)=\mu_{h2} \end{array} \right ..
\label{eq66}
\end{eqnarray}
Hence $\mu(t_1) = \mu(t_1-1)$, the minority action is to stay in the same node and gives the positive payoff to the winning strategy
\begin{eqnarray}
R_{\alpha_n^s}(t_1-1)=-a_{\alpha_n^s} A(t_1-1).
\label{eq67}
\end{eqnarray}
\item[\underline{Stage 3}]
~\\
There is a non-zero probability that strategies corresponding to the first $l/2$ utilities in $\{u_l\}$ have won in the last step. 
Such circumstance is possible provided stages 1 and 2 are realized.
If this third condition is fulfilled then we mark such history $\mu_C$. 
Then all first $l/2$ strategies suggest the same reaction after $\mu_C$. 
Hence the majority decision at $t_1$ is to stay in the node and the maximal demand (\ref{eq65}) is generated.
All strategies with high utility get the penalty and the low-utility ones are rewarded by the same amount. 
The game follows the minority decision and escapes from the $\mu_C$ de
Bruijn node. When the game leaves $\mu_C$, the strategy set is split into
two groups of high and low utility, as illustrated in Fig.~\ref{fig62}. In
the next steps the game goes to $\mu\neq\mu_C$.
\begin{figure}[h]
\begin{center}
\begin{tabular}{cc}
\includegraphics[scale=.4]{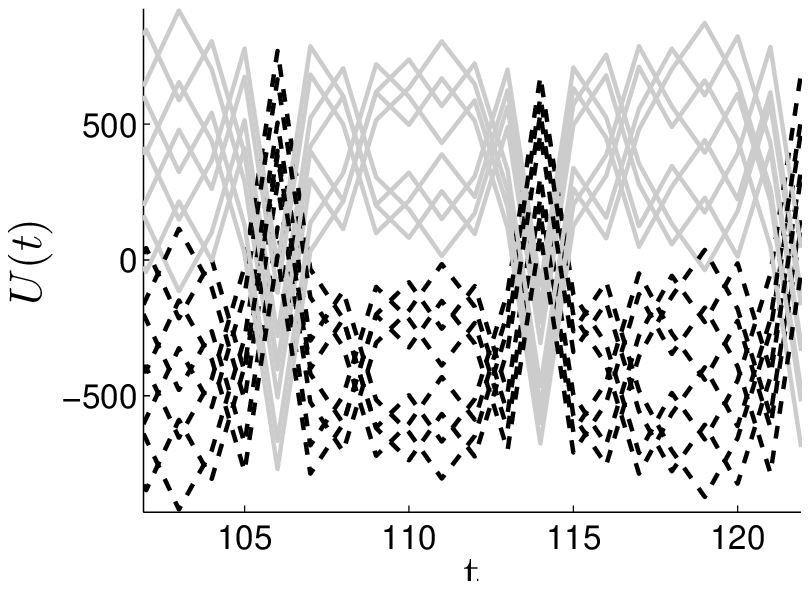} &
\includegraphics[scale=.35]{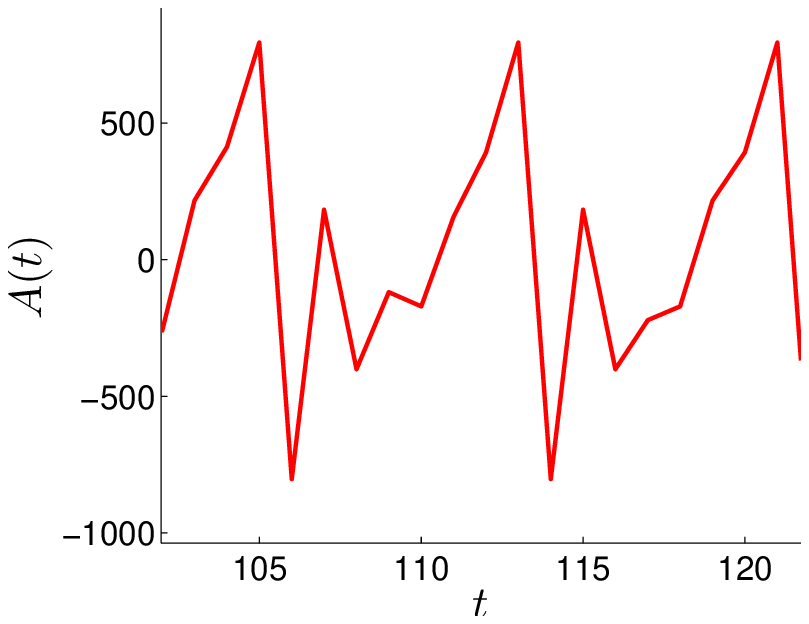}
\end{tabular}
\end{center}
\vspace*{8pt}
\caption{\label{fig62}The time evolution of the utilities (left) and the
aggregated demand (right) for the MG with $N=1601$, $S=2$, $m=2$ and
$g(x)=x$.}
\end{figure}
\item[\underline{Stage 4}]
~\\
Next steps do not substantially affect utilities as long as the
history $\mu_C$ does not reappear. There is no history other than $\mu_C$
assuring that the first $l/2$ strategies in the $\{u_l\}$ list suggest a
collective action resulting with the most spiky demand. 
Hence, after $t_1$, the variations of $A$ do not affect the utility significantly
untill the $\mu_C$ reappears at $t_2>t_1$ when the set of the best $l/2$
strategies is the same as at $t_1$. Then the $l/2$ best strategies suggest
the game to shift to another node characterized by history $\mu(t_2+ 1)
\neq \mu_C$ and the maximal demand $|A(t_2)| = N(1-\frac{1}{2^{S-1}})$ is
generated. All the $l/2$ best strategies get penalty proportional to
the absolute value of the aggregated demand. Concurrently, the $l/2$ strategies
with the lowest utility are rewarded with the same amount (cf.
Fig.~\ref{fig62}).
\item[\underline{Stage 5}]
~\\
Next, the game follows the edge leading to the same node. Subsequently,
the $l/2$ best strategies suggest staying in the same vertex $\mu_C$.
Again, high absolute value of demand is generated but the sign of
$A(t_2+1)$ is opposite to the sign of $A(t_2)$. Consequently, all
strategies with high $U(t_2 + 1)$ get penalty $N(1-\frac{1}{2^{S-1}})$
and, concurrently, strategies with low utility get reward of the same size.
\item[\underline{Stage 6}]
~\\
The game goes to the vertex $\mu_C(t_2 + 2) \neq \mu_C$ and the scenario from stages 4--6 repeats.
\end{itemize}

\section{Symbol captions}

\vspace{5pt}

\begin{tabular}{ll}
$a_\alpha$ & -- action suggested by strategy $\alpha$\\
$a^\ast$ & -- the minority action \\
$\alpha_n^s$ & -- the $s$-th strategy of the $n$-th agent \\
$\alpha_n^\prime$ & -- the active strategy, or the strategy of the highest utility, \\ & for the $n$-th agent \\
$\alpha^B$ & -- the best strategy from the whole set of strategies in the game \\
$A=\sum_{n=1}^Na_{\alpha_n^\prime}$ & -- aggregated demand \\
${\mathbb E}\,A(x_i)$ & -- expected value of demand over possible realizations of the game \\
$\{\beta_i\}_{i=1}^{2^P}$ & -- set of $2^P$ pairwise different strategies \\
$\Delta_n$ & -- set of $S$ strategies of the $n$-th agent \\
$f$ & -- frequency of demand peaks \\
$g$ & -- payoff function \\
$m$ & -- length of the sequence of last minority decisions \\
$\mu=[a^\ast(t-m),\ldots,a^\ast(t-1)]$ & -- sequence of the last minority decisions \\
$\mu_C$ & -- history of minority decisions preceding first strong fluctuation of demand \\
$\mu_{h1(2)}=[\,-(+)1,\ldots,-(+)1\,]$ & -- homogenous de Bruijn nodes \\
$N$ & -- the total number of agents in the game \\
$P=2^m$ & -- number of possible realizations of $\mu$ \\
${\mathcal P}_{min(max)}$ & -- probability that the minimal (maximal) utility of any agent attains \\ & the absolute minimum (maximum) value $U_{min(max)}$ \\
$\rho(n)$ & -- distribution of strategies for the $n$-th agent at the beginning of the game \\ 
$R_\alpha$ & -- payoff for the strategy $\alpha$ \\
$S$ & -- the total number of strategies for each agent \\
$\{u_l\}_{l=1}^{2^m+1}$ & -- ordered list of different utility values when the extremal value of \\ & $U_{min(max)}$ is attained \\
$U_\alpha$ & -- utility of the strategy $\alpha$ \\
$U_{min(max)}=-(+)2^m$ & -- the absolute minimum (maximum) value of the utility \\
$x(t)=[\,\mu(t),\,U_1(t),\ldots,U_{2^P}(t)\,]$ & -- state of the game at time $t$ \\
\end{tabular}

\end{document}